\documentclass[aip,preprint,floatfix]{revtex4-1}
\usepackage[version=3]{mhchem}
\usepackage{graphicx}
\usepackage{amsmath}
\usepackage{amssymb} 
\usepackage{mhchem}
\usepackage{color}
\usepackage[toc,page]{appendix}
\newcommand{\cF}{\mathcal{F}}
\newcommand{\mr}{\mathbf{r}}
\newcommand{\rid}{\mathrm{id}}
\newcommand{\rex}{\mathrm{ex}}
\newcommand{\rcoul}{\mathrm{coul}}
\newcommand{\rhc}{\mathrm{hc}}
\newcommand{\rsr}{\mathrm{sr}}
\newcommand{\rres}{\mathrm{res}}
\newcommand{\rin}{\mathrm{in}}

\begin{document} 

\title{ Charge neutrality breakdown in confined aqueous electrolytes: theory and simulation}

\author{Thiago Colla}
\email{thiago.colla@ufrgs.br}
\affiliation{Instituto de F\'isica, Universidade Federal do Rio Grande do Sul, Caixa Postal 15051, CEP 91501-970, Porto Alegre, RS, Brazil}

\author{Matheus Girotto}
\email{matheus.girotto@ufrgs.br}
\affiliation{Instituto de F\'isica, Universidade Federal do Rio Grande do Sul, Caixa Postal 15051, CEP 91501-970, Porto Alegre, RS, Brazil}

\author{Alexandre P. dos Santos}
\email{alexandre.pereira@ufrgs.br}
\affiliation{Instituto de F\'isica, Universidade Federal do Rio Grande do Sul, Caixa Postal 15051, CEP 91501-970, Porto Alegre, RS, Brazil}

\author{Yan Levin}
\email{levin@if.ufrgs.br}
\affiliation{Instituto de F\'isica, Universidade Federal do Rio Grande do Sul, Caixa Postal 15051, CEP 91501-970, Porto Alegre, RS, Brazil}

\begin{abstract}

We study, using Density Functional theory and Monte Carlo simulations, 
aqueous electrolyte solutions between charged infinite planar surfaces, in a contact with a bulk salt reservoir. 
In agreement with recent experimental observations [Z. Luo {\it et al.}, Nat. Comm. {\bf 6}, 6358 (2015)], we find that
the confined electrolyte lacks local charge neutrality. 
We show that a Density Functional Theory (DFT) based on a bulk-HNC expansion properly accounts for strong electrostatic correlations and allows us to accurately calculate the ionic density profiles between the charged surfaces, even for electrolytes containing trivalent counterions. The DFT allows us to 
explore the degree of  local charge neutrality violation, as a function of plate separation and bulk electrolyte concentration, and 
to accurately calculate the interaction force between the charged surfaces.

\end{abstract}

\maketitle

\section{Introduction}

In physics,  as well as in chemistry, one often finds situations in which electrolyte 
is confined between charged surfaces. 
These surfaces may belong to electrodes, macromolecules, charged colloidal particles, polymers, etc., and can 
give rise to complex ionic distributions generally known as Electrical Double Layers (EDLs).  
Presence of electrolyte between charged surfaces strongly affects their interaction and can lead to 
fascinating phenomena such as like-charge attraction~\cite{Le02,Lob99,Sam11} and charge reversal~\cite{Le05,Vin05}. The interaction between EDLs is fundamental for understanding colloidal stability and efficient energy storage\cite{Le02,Lik01}. Electrolytes confined by porous walls show promising application as supercapacitors, since carbon-based electrodes are known to increase the storage performance of these devices due to their high specific surface area \cite{Zha09}. The porosity brings about strong ionic confinement within the nano-sized pores. Despite intense exploration, many questions are yet to be elucidated regarding the behavior of strongly correlated Coulomb systems in a confined environment \cite{Zha09}.

From a theoretical perspective, the earliest attempts to describe the properties of electrolytes in a close vicinity of charged surfaces go  back to the pioneering work of Gouy, Chapman, and Stern (GCS) \cite{Gou10,Cha13,Ste24}. In the original approach, the double layer structure was described as a thin layer of counterions,  condensed onto a charged surface -- the so-called Stern-layer -- followed by a diffuse region in which ionic distributions rapidly decayed to their bulk values \cite{Zha09}. This simple picture was based on the mean-field Poisson-Boltzmann (PB) theory. It provided a clear physical explanation for a number of interesting phenomena inherent to electrical double layers -- ranging from the charge renormalization of charged surfaces \cite{Le02,Tri02} to the capacitance of simple electrodes \cite{Fed14,Zha09}. For this reason the GCS theory has remained very popular in both electrochemistry and biophysics literature \cite{Fed14}. Care, however, is required when extrapolating this simple 
picture to the case of strongly correlated electrolytes -- such as aqueous electrolytes made of multivalent ions, organic electrolytes, or ionic liquids. In these cases, the structure of the EDL is dominated by ionic correlations \cite{Pai11}, giving rise to complex phenomena such as charge reversal in multivalent ionic electrolytes and layering in ionic liquids \cite{Wu11}. Even in situations of low electrostatic couplings, the packing effects in strongly confined electrolytes can alone be responsible for oscillatory profiles which cannot be captured at a mean-field level of GCS theory \cite{Fry12}.  

Over the last decades, a number of different approaches have been put forward to 
quantitatively describe the ionic correlations in EDLs beyond the 
mean-field theory \cite{Att96}. A great progress has been achieved with computer simulations, 
in part  due to rapidly increasing computational power,  but  also because of the development of new 
techniques which allow to efficiently handle the long-range Coulomb interactions  in
slab geometry \cite{Ber99,Arn02,San16}. 
The theoretical advances  have relied on  extending the integral equations (IE) methods developed 
for bulk systems \cite{Hen78,Hen79,Ver82,Ball86} to inhomogeneous ones \cite{Kje86,Kje88,Kje88_2,Kje92,Gre94}, as well as to interacting EDLs \cite{Loz84,Loz86,Loz90,Ale90}. 
Other approaches explored Modified Poisson-Boltzmann (MPB) equation \cite{Out11,Bhu07,Sta09}, and a classical Density Functional Theory (DFT) \cite{Jai14}.  The main advantage of using DFT is that various contributions to free energy can be handled separately, allowing distinct approximations in their calculations. For instance, it is well known that the excluded volume interactions can be described to a very high degree of accuracy using the Fundamental Measure Theory (FMT) \cite{Ros89,Ros90,Ros02,Roth10}, and can be conveniently decoupled from the electrostatic interactions. The DFT can be constructed using relatively simple arguments based on truncation of functional expansions, introduction of coupling parameters, local or weighted density approximations, or even combinations of these \cite{Jai14,Yan15}. All these approximations are easy to control and possess quite transparent physical interpretations, allowing for their systematic improvement. This should be contrasted with IE techniques, for which the 
approximations  are based on closure relations
which appear in diagrammatic cluster expansions for the correlation and bridge functions \cite{Att96,Loz92}, the degree of accuracy of which is, in general, not known {\it a priori}.

It is usual to take for granted the local charge neutrality between charged surfaces~\cite{DoLe15,DiLe15,Whi93} when dealing with EDL in both theoretical and computational approaches. This is not to be confused with the \textit{overall} charge neutrality, which clearly has to be always satisfied for a charged system as whole. In order to avoid any confusion, we emphasize that from now on the term \textit{charge neutrality} will refer to the \textit{local} electroneutrality of electrolyte confined between charged walls. The possibility that electrolytes in contact with a bulk reservoir might lack charge neutrality in situations of strong confinement has been already discussed both theoretically \cite{Loz96_1,Loz96_2} and observed in computer simulations \cite{Lo98,Lee97,Lee99}. However, these findings have only very recently been confirmed experimentally in the work of Luo \textit{et al.}~\cite{Wu15}, who demonstrated that electrolytes confined in narrow pores do not in general exhibit electroneutrality.
To explore this issue  Luo \textit{et al.} used Nuclear Magnetic Resonance (NMR) techniques to demonstrate that electrolytes confined by nanoporous carbon with graphite-like internal surfaces can violate charge neutrality. They have observed a dependence of charge neutrality breakdown on ion specificity which followed the Hofmeister series.  This is consistent with the recent theoretical works which show that
ionic specificity arises from a combination of hydrophobic, dispersion, and polarization interactions, 
which are very different for  
$F^-$, $Cl^-$, $Br^-$, and $I^-$ near hydrophobic surfaces~\cite{Le09,Le11,LeDo14,To06,To02,Gho05,To01,Nim04}.
Although this ionic specificity is extremely interesting, we will not address it in this paper. 
Instead we will explore the role that electrostatic correlations~\cite{Le02} play in charge neutrality breakdown. A similar line of investigation has been taken by Lozada-Cassou and co-workers \cite{Loz96_1,Loz96_2}, who used a three-point extended version of the traditional IE formalism to address the problem of electroneutrality violation in electrolytes confined by parallel charged surfaces. In this approach, the two-plate system was modeled as ``dumbbell-like'' molecules at infinite dilution, and the resulting ionic profiles were calculated as thee-body wall-wall-ion correlations \cite{Loz84,Loz86,Loz90,Loz90_2,Ale90,Fel93,Jim09}. The ionic correlations were treated altogether at the level of the Mean Spherical Approximation (MSA). It is, however, well known that in situations in which electrolyte is bounded by narrow pores --  where violation of electroneutrality is expected to take place -- size correlations between the confined ions may strongly influence the structure of the EDLs \cite{Zhou89,Hen98_1,Hen98_2,Kam08}. Furthermore, electrostatic correlations in electrolytes containing multivalent ions usually require use of more accurate approximations. It is, therefore, unclear to what extend can strong correlations be captured at the MSA level. In order to further explore the question of how important are the correlation effects in determining electroneutrality violation we shall, therefore, apply a somewhat different approach. The size correlations will be described through the FMT approach, whereas the electrostatic correlations will be taken into account at the level of the hyppernetted chain (HNC) approximation. Furthermore, a recently developed MC simulation technique, well suited to simulate ionic systems in slab geometries at very low computational cost will be applied \cite{San16} to this system. 
We will show that for large separations between surfaces 
the macroscopic charge neutrality is restored and the traditional Donnan approach, in which a potential difference across the system-reservoir interface is established to force the overall electroneutrality (Donnan equilibrium) \cite{Don24,Ohshi85}, can be applied.  On the other hand, for narrow pores the interior of the pore is not electroneutral, in which case the internal charge of the pore will be determined by the chemical equilibrium with a bulk reservoir. These findings are in qualitative agreement with earlier theoretical predictions obtained in Refs. [\onlinecite{Loz96_1}] and [\onlinecite{Loz96_2}]. We will also investigate to what extent the absence of electroneutrality between the confining plates influences the net forces between the charged surfaces. It is important to emphasize that such effects from local violation of electroneutrality on the interaction between double layers can have important implications in a number of biological and physical systems composed by charged objects immersed in an electrolyte solution \cite{Le02}. Examples range from colloidal self-assembly \cite{Odri13,Odri16,Fur15,Cade15} to membranes \cite{Jim04} and biological cells \cite{Gel00}, as well as confined phase separations \cite{Loz96},  ionic correlations across charged membranes \cite{Loz97} and colloidal systems confined by charged walls \cite{Jim09}. For a recent overview on the way EDL interaction might influence the assembly of soft-matter systems, we refer the reader to Ref. [\onlinecite{editor}] and references therein. In spite of these numerous applications, the role that the charge neutrality violation plays in the  EDL interactions has been mostly neglected after the original contributions of Lozada-Cassou {\it et al.} \cite{Loz84} which addressed this issue. The aim of the present work is to investigate the effects that addition of salt and electrostatic correlations have on the charge neutrality violation, and explore how it influences the interaction between planar EDLs.

The work is organized as follows. In Sec. II, a description of the system under consideration is briefly given. Some details about the MC techniques and the DFT are presented in Secs. III and IV, respectively. 
Results are shown in Sec. V
and the Conclusions are outlined in Sec. VI. Technical details regarding some of the specific calculations employed in this work can be found in the Appendices.   

\section{Model System}

%%%%%%%%%%%%%%%%%%%%%%%%%%% Fig. 1 %%%%%%%%%%%%%%%%%%%%%%%%%%%%%%%%%
\begin{figure}
\centering
\includegraphics[width=16cm,height=5.5cm]{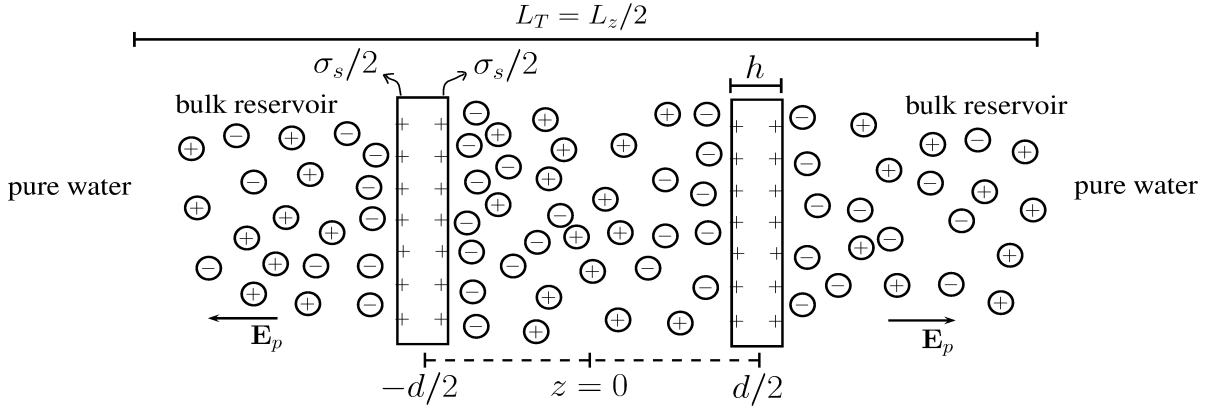}
\caption{Schematic representation of the system. The positively charged walls of thickness $h \rightarrow 0$ are located at $z=\pm d/2$, with the confined electrolyte between them. As indicated in the figure, each face of a surface carries a charge density $\sigma_s/2$.  
%The confined electrolyte is in chemical equilibrium with the bulk salt reservoir located beyond the walls. 
The system consists of a confined electrolyte and a bulk solution, which are in chemical equilibrium and can freely exchange ions. The net charge in the region $-d/2<z<d/2$,  $Q_t$, is determined by the chemical equilibrium with the bulk. In order to implement 3d Ewald summation for 2d slab geometry, in the simulations the region  $2L_T>|z|>L_T$ is occupied by pure solvent.}
\label{fig:fig1}
\end{figure}
%%%%%%%%%%%%%%%%%%%%%%%%%%%%%%%%%%%%%%%%%%%%%%%%%%%%%%%%%%%%%%%%%%%%%

Throughout the paper we will use the framework of the Primitive Model (PM) of electrolytes, in which the ions are represented as hard spheres with point charge embedded at the center. The solvent is modeled as a uniform dielectric continuum with its molecular structure disregarded. We will consider a system composed of two positively charged walls located along the $z$ direction at positions $z=-d/2$ and $z=d/2$. Both walls have a surface charge density $\sigma_s$ uniformly distributed over their infinite surfaces. The space between the walls is occupied by an aqueous electrolyte (permittivity $\epsilon=80\epsilon_0,$ with $\epsilon_0$ the dielectric constant of vacuum) composed of cations of valence $\alpha_{+}$ and anions of valence $\alpha_{-}$. The confined electrolyte is in contact with a bulk reservoir of concentration $c_{s}$. % and $\alpha_{+}c_{s}/\alpha_{-}$ for cations and anions, respectively.
In simulations, we will consider that the whole system (electrolyte plus reservoir) is confined by neutral walls symmetrically located at positions $\pm L_{T}/2$, such that $L_{T}\gg d$ (see Fig. 1). Due to the symmetry in the $XY$ plane, all the system inhomogeneity takes place along the $z$ direction. Although the net charge in the space between the walls is not explicitly taken to be zero, it is important to point out that the system as a whole does obey the global charge neutrality, which is expressed as:
\begin{equation}
2\sigma_s+\sum_{i}q_{i}\int_{-L_T/2}^{L_T/2} \rho_{i}(z)dz=0,
\label{total_cn}
\end{equation}  
where $\rho_{i}(z)$ represents the ionic profile of component $i=\pm$ at position $z$, and $q_i=\alpha_iq$ is its corresponding charge (with $q$ being the charge of a proton).

\section{Model and Monte Carlo Simulations}

The difficulty of simulating systems with long-range interactions is that the interaction potential can not be cutoff and one can not use simple periodic boundary conditions.  Instead one must consider an infinite set of replicas of the simulation cell, so that the ions in the principal cell feel
the electric field produced by the ions from all the replicas. To efficiently sum over the replicas one must use some form of Ewald summation~\cite{Fr02}. There are a number of very efficient implementations of Ewald summation for isotropic 3d Coulomb systems. The situation,
however, is more complicated for systems with reduced symmetry, such as our two infinite charged plates,  
in which case the system is periodic only in two out of three dimensions. 
Recently we have introduced a method which allows for a very efficient adaptation of 3d Ewald 
summation to systems with reduced slab geometry, with confining charged surfaces. The fundamental idea of the method 
is to separate the electric field produced by the infinite charged 
plates from the rest of the system.  The
difficulty, however, is that when this is done, the residual systems is no longer charge neutral,
so that the electrostatic energy of the infinitely replicated system will diverge. We showed, however, that this divergence can be renormalized away, allowing us to calculate finite renormalized electrostatic energy.  The details of all the derivations can be found in Ref. [\onlinecite{San16}]. Here we will
just present the applications of the algorithm to the present problem.

Our simulation cell has the volume $V=L_xL_yL_z$, with $L_x=L_y=L=50$~\AA\ and $L_z=250$~\AA. The
electrolyte is in the region $-L/2 < x < L/2, -L/2 < y < L/2, -L_T/2 < z < L_T/2$ (with $L_T=L_z/2$), and the pair of charged plates are positioned at $-d/2$ and $d/2$. In the regions $-L_z/2 < z < -L_T/2$ and $L_T/2 < z < L_z/2$ there is pure water, see Fig.~\ref{fig:fig1}.
Water is treated as a continuum of dielectric constant $\epsilon_w=80\epsilon_0$. The Bjerrum length, defined as $\lambda_B=q^2/\epsilon_w k_B T$, is  $7.2~$\AA. This value is appropriate for water at room temperature. All ions have radius $2~$\AA . 
The plates are equally charged with a positive surface charge density in the range $0.05~\mathrm{C/m^2}<\sigma_s<0.07~\mathrm{C/m^2}$. The effect of counterion valence  is explored by changing between 1:1, 2:1, and 3:1 electrolyte.  
 
To efficiently sum over all the replicas the electrostatic potential is split into long and short-range contributions, in addition to the potential produced by the charged surfaces, see Ref [\onlinecite{San16}], and a residual potential of self-interaction.  %The electrostatic energy
%can then be written as
%\begin{equation}
%U=U_S+U_L+U_{self}+U_{cor}+U_p \ .
%\end{equation}
The short-range electrostatic potential at position ${\pmb r}$ is
%%%%%%%%%%%%%%%%%%%%%%%
\begin{equation}\label{phi_short}
\phi_S({\pmb r})=\sum_{j=1}^{N} q_j\frac{\text{erfc}{(\kappa_e |{\pmb
r}-{\pmb r}_j|)}}{\epsilon_w |{\pmb r}-{\pmb r}_j|} \ ,
\end{equation}
%%%%%%%%%%%%%%%%%%%%%%%%%%%%%%
The dumping parameter $\kappa_e$ is set to $\kappa_e=5/L$.
The long-range electrostatic potential is
%%%%%%%%%%%%%%%%%%%%%%%%%%%%%%
%\begin{eqnarray}\label{phi_long1}
%\phi_L({\pmb r}) = \sum_{{\pmb k}}\frac{4\pi}{\epsilon_w V |{\pmb k}|^2}
%\text{exp}\left(-\frac{|{\pmb k}|^2}{4\kappa_e^2}\right)\sum_{j=1}^{N}q_j\text{exp}[i{\pmb k}\cdot({\pmb r}-{\pmb r}_j)] \ .
%\end{eqnarray}
%%%%%%%%%%%%%%%%%%%%%%%%%%%%%%
%%%%%%%%%%%%%%%%%%%%%%%%%%%%%%
\begin{eqnarray}\label{phi_long2}
\phi_L({\pmb r}) = \sum_{{\pmb k}\neq{\pmb 0}}\frac{4\pi}{\epsilon_w V |{\pmb k}|^2}
\text{exp}\left(-\frac{|{\pmb k}|^2}{4\kappa_e^2}\right)\sum_{j=1}^{N}q_j\text{exp}[i{\pmb k}\cdot({\pmb r}-{\pmb r}_j)] - \frac{2\pi}{\epsilon_wV}\sum_{j=1}^Nq_j(z-z_j)^2 \ .
\end{eqnarray}
%%%%%%%%%%%%%%%%%%%%%%%%%%%%%%
The number of ${\pmb k}$-vectors defined as ${\pmb k}=(2\pi n_x/L,2\pi n_y/L,2\pi n_z/L_z)$, where $n's$ are integers, is set to around $400$ in order to achieve fast convergence.
The electrostatic potential produced by the charged surfaces is 
%%%%%%%%%%%%%%%%%%%%%%%%%%%%%%
\begin{align}
\phi_p^{\rcoul}({\pmb r})= \begin{cases}
\dfrac{4\pi}{\epsilon_w}\sigma_s(z+d/2) & z<-d/2,\\
0 & -d/2<z<d/2,\\
-\dfrac{4\pi}{\epsilon_w}\sigma_s(z-d/2) & d/2<z.
\end{cases} 
\label{elec}
\end{align}
%%%%%%%%%%%%%%%%%%%%%%%%%%%%%%
The self-interaction potential has the form 
\begin{equation}
\phi^{\mathrm{self}}({\pmb r})= q_i\frac{\text{erf}{(\kappa_e |{\pmb
r}-{\pmb r}_i|)}}{\epsilon_w |{\pmb r}-{\pmb r}_i|} \ ,
\end{equation}
We can now easily compute the total ionic electrostatic energy
%%%%%%%%%%%%%%%%%%%%%%%%%%%%%%
\begin{equation}\label{energy}
U=\frac{1}{2}\sum_{i=1}^Nq_i[\phi_L({\pmb r}_i)-\phi^{\mathrm{self}}({\pmb r}_i)] + \frac{1}{2}\sum_{i\neq j}^Nq_i\phi_S({\pmb r}_i) + \sum_{i=1}^Nq_i\phi_p^{\rcoul}({\pmb r}_i) \ .
\end{equation}

To perform MC simulations we use Metropolis algorithm with $10^6$ MC steps to achieve equilibrium. The profile and force averages are performed with $5\times10^5$ uncorrelated samples. During the equilibration, we adjusted the length of the particle displacement to achieve an acceptance of trial moves near $50\%$. We are particularly interested in the net charge of electrolyte in between the confining surfaces.  The external electrolyte acts as a reservoir for the internal region, so that the ions are allowed to freely move across the charged surfaces.
The interaction between two surfaces is modulated by the external and internal electrolyte, and has both  electrostatic and entropic contributions. 
To calculate the mean electrostatic force we use the method
of virtual displacement in which one of the plates is moved
while the other plate and all the ions remain fixed, which implies that the electrostatic force per unit area in the $z$ direction is
\begin{equation}
\langle  F_z^{\rcoul} \rangle =\dfrac{2\pi}{\epsilon_w}\sigma_{s}^2-\frac{1}{A}\sum_{j=1}^N \left\langle\frac{\partial U_p({\pmb r}_1,...,{\pmb r}_N)}{\partial z_j} \right\rangle,\
\label{Fz}
\end{equation}
where $A$ is the area of the plate, $U_p=\sum_{i=1}^Nq_i\phi^{\rcoul}_{p1}({\pmb r}_i)$, and  
\begin{align}
\phi^{\rcoul}_{p1}({\pmb r}) = \begin{cases}
\dfrac{2\pi}{\epsilon_w}\sigma_s(z+d/2) & z<-d/2,\vspace{0.5cm}\\
-\dfrac{2\pi}{\epsilon_w}\sigma_s(z+d/2) & -d/2<z.
\end{cases} 
\label{elec}
\end{align}
The first term on the right-hand side of Eq. (\ref{Fz}) is the mutual force between the charged wall surfaces, while the second term represents the ionic-averaged electrostatic forces on the first wall. Note that a positive force means repulsion between the walls. Eq. (\ref{Fz}) can be simplified to yield,
\begin{equation}
\langle  F_z^{\rcoul} \rangle =\dfrac{2\pi}{\epsilon_w}\sigma_{s} (\sigma_>-\sigma_<)\,,
\end{equation}
where $\sigma_{>}$ and $\sigma_{<}$ represent the total charge per unit of area located in the regions $z>-d/2$ and $z<-d/2$, respectively.

To calculate the entropic force, which arises from the transfer of momentum in the collisions between ions and plates, we use the method introduced by Wu~{\it et al.}~\cite{WuBr99} It consists of performing a small virtual displacement of the plates along the $z$ direction -- while all the ions remain fixed -- and counting the number of resulting  virtual overlaps between the plates and the ions. The entropic force per unit area can then be written as
\begin{equation}
\beta F_z^{\mathrm{en}}=\frac{<N^c>-<N^f>}{2\Delta R A} ,
\end{equation}
where $N^c$ is the number of virtual overlaps between the plates with the ions after a small displacement $\Delta R=0.9~$\AA~ that brings plates closer together (superscript $c$ stands for closer) and $N_f$ is the number of overlaps of the plates with the ions after a displacement $\Delta R$ that moves the two plates farther apart (superscript $f$ stands for farther)~\cite{Le12}.

The profiles were made counting the average number of particles in a volume range $\text{bin}=\Delta zL^2$ running over the $z$ direction. The value of $\Delta z$ was set to $0.5$~\AA.

% To make empty space larger than the periodic dimensions was proved useless by a series of tests performed varying $L_z$, as all tests converged to the same equilibrium state. This result is contrary to the suggestion of previous work by Spohr\cite{Sp97}, but we believe this is due to the total charge neutrality, $\sum_i^Nq_i=0$, of the system, therefore making the unit cell weakly interacting with the next ones in the $z$ direction.

\section{Density Functional Theory}

The DFT assumes that the functional of a 
set of densities $\{\rho_{i}(\mr)\}$, 
\begin{equation}
\Omega[\{\rho_{i}(\mr)\}]=\cF[\{\rho_{i}(\mr)\}]+\sum_{i}\int[\mu_{i}+\phi_{i}(\mr)]\rho_{i}(\mr)d\mr, 
\label{grand}
\end{equation}
where $\mu_{i}$ and $\phi_{i}(\mr)$ are respectively the chemical potential and the external potential acting on the ion of type $i$, achieves a minimum at equilibrium \cite{Evans92}.  This minimum corresponds to the system's grand potential (or the ground-state energy of a quantum system \cite{Merm65}). The functional  $\cF[\{\rho_i(\mr)\}]$ is the {\it intrinsic free energy}, since it contains all the information about the particle interactions, regardless of any particular external potential acting upon the system. It can be split into an ideal gas contribution $\cF^{\rid}$ plus an excess free energy $\cF^{\rex}$, in which all contributions resulting from the particle interactions are included \cite{Evans79,Singh91,Low02,Evans92,Wu07}. The ideal gas contribution is:
\begin{equation}
\beta \cF^{\rid}[\{\rho_i(\mr)\}]=\sum_{i}\int{\rho_{i}(\mr)[\ln(\Lambda^3 \rho_{i}(\mr))-1]d\mr},
\label{ideal}
\end{equation}
where $\Lambda$ is the usual de Broglie wavelength. Use of this exact relation together with a straightforward application of the Euler-Lagrange minimization condition to Eq.~(\ref{grand}) provides the following equilibrium distributions:
\begin{equation}
\rho_{i}(\mr)=\bar{\rho}_{i}\exp(-\beta\phi_i(\mr)+c_i(\mr)),
\label{dist1}
\end{equation}
where $\bar{\rho}_{i}=\exp(-\beta \mu_i)/\Lambda^3$, and $c_{i}(\mr)=-\beta\delta\cF^{\rex}/\delta\rho_i(\mr)$ defines the first-order direct correlation function. If the density profiles corresponding to a given external potential $\phi^{0}_{i}(\mr)$ are known, the equilibrium distributions (\ref{dist1}) can be easily written in terms of such reference profiles as:
\begin{equation}
\rho_{i}(\mr)={\rho}^{0}_{i}(\mr)\exp(-\beta\delta\phi_i(\mr)+\delta c_i(\mr)),
\label{dist2}
\end{equation}
where $\rho^{0}_{i}(\mr)=\bar{\rho}_{i}\exp(-\beta\phi^{0}_{i}(\mr)+ c^{0}_{i}(\mr))$ are the reference density profiles, with $\delta \phi_{i}(\mr)=\phi_i(\mr)-\phi^{0}_i(\mr)$ and $\delta c_i(\mr)=c_{i}(\mr)-c^{0}_i(\mr)$. Unfortunately, the intrinsic energy, $\cF^{\rex}[\{\rho_{i}(\mr)\}]$,  is not known for arbitrary density distributions.  However, it can be very accurately calculated, for example, for a bulk system which can then be used as the reference state.  In general, if a suitable approximation for the reference fluid with densities $\rho_{i}^{0}(\mr)$ is known, Eq. (\ref{dist2}) can then be applied in a perturbative scheme to obtain improved corrections for the desired distributions $\rho_{i}(\mathbf{r})$, as we will see later on.

So far no approximations have been made in writing Eqs. (\ref{dist1}) and (\ref{dist2}). The excess free energy can be further split in accordance with different particle interactions. In the present case of ionic systems in the framework of the PM approach, the functional $\cF^{\rex}$ is a combination of hard-core $\cF^{\rhc}$ and electrostatic $\cF^{\rcoul}$ contributions, $\cF^{\rex}=\cF^{\rhc}+\cF^{\rcoul}$. Accordingly, the single-particle direct correlation function can be written as a sum of hard-core and electrostatic contributions, $\delta c_{i}(\mr)=\delta c_{i}^{\rhc}(\mr)+\delta c_{i}^{\rcoul}(\mr)$. This possibility of treating separately the different contributions for the particle correlations is one of the major advantages of using a DFT approach. The hard-core contribution can, for example, be treated in the framework of the Fundamental Measure Theory (FMT) \cite{Ros89,Ros90,Ros02}, which has proven to be extremely accurate in describing both thermodynamic and structural properties of hard-
sphere system for a variety of packing fractions and size asymmetries \cite{Roth10,Roth16}. We will, therefore, apply the FMT for calculating the ionic exclusion volume interactions, using its more recent White-Bear formulation \cite{Roth02,Yu02}, the details of which are outlined in Appendix A. For a deeper analysis of the hard-sphere FMT we refer the reader to Roth's recent review on this topic, Ref. [\onlinecite{Roth10}]. 

The difference between the electrostatic single-particle direct correlation function and the reference state, $\delta c_{i}^{\rcoul}(\mr)$, can be formally evaluated by introducting a parameter $\lambda$ to continuously interpolate between the equilibrium and reference states. The simplest choice is to use a linear path, such that $\rho_{i}^{\lambda}(\mr)=\lambda\delta\rho_{i}(\mr)+\rho_{i}^{0}(\mr)$, with $0\le\lambda\le1$ and 
$\delta\rho_{i}(\mr)\equiv\rho_{i}(\mr)-\rho_{i}^{0}(\mr)$. When $\lambda$ varies from zero to unity, the profiles $\rho_{i}^{\lambda}(\mr)$ smoothly change between the reference state and the desired equilibrium distributions. The corresponding variation in the single-particle electrostatic correlations with respect to their reference counterpart can be written as:
\begin{equation}
\delta c_{i}^{\rcoul}(\mathbf{r})\equiv c_{i}^{\rcoul}(\mr;\lambda=1)-c_{i}^{\rcoul}(\mr;\lambda=0)=\int_{0}^{1}\dfrac{\partial c_{i}^{\rcoul}(\mr;\lambda)}{\partial\lambda}d\lambda,
\label{dc1}
\end{equation}
where $c_{i}(\mr;\lambda)$ represent the direct correlation functions evaluated when the profiles are $\rho_{i}^{\lambda}(\mr)$. The derivative on the right-hand side of the last equality can be obtained by noting that the single-particle correlations depend on $\lambda$ only implicitly, through the density profiles $\rho_{i}^{\lambda}(\mr)$. It can, therefore, be related to the changes in the density distributions as \cite{Hansen}:
\begin{equation}
\dfrac{\partial c_{i}^{\rcoul}(\mr;\lambda)}{\partial\lambda}=\sum_{j}\int\dfrac{\delta c_{i}^{\rcoul }(\mr; \lambda)}{\delta \rho_{j}^{\lambda}(\mr')}\dfrac{\partial\rho_{j}^{\lambda}(\mr')}{\partial\lambda}d\mr'=\sum_{j}\int c_{ij}^{\rcoul}(\mr,\mr';\lambda)\dfrac{\partial\rho_{j}^{\lambda}(\mr')}{\partial\lambda}d\mr',
\label{dc2}
\end{equation}
where in the second equality the definition of the direct pair correlation function has been used. 
%The first equality follows from a linear functional expansion for small variations of $\Delta\rho_{i}^{\lambda}(\mr)\equiv\rho_{i}^{\lambda+\Delta\lambda}(\mr)-\rho_{i}^{\lambda}(\mr)$, which than becomes exact after taking the limit when $\Delta\lambda$ goes to zero \cite{Hansen}. 
With the particular choice of the linear dependence of $\rho_{i}^{\lambda}(\mr)$ on $\lambda$, the derivative on the right-hand side of Eq. (\ref{dc2}) reduces to  $\delta\rho_{j}(\mr')$, and the changes in the direct correlation functions described in Eq. (\ref{dc1}) become:
\begin{equation}
\delta c_{i}^{\rcoul}(\mr)=\sum_{j}\int\delta\rho_{j}(\mr')\bar{c}^{\rcoul}_{ij}(\mr,\mr')d\mr'.
\label{ci1}
\end{equation}   
In the above relation, we have defined the mean pair direct correlation function as:
\begin{equation}
\bar{c}^{\rcoul}_{ij}(\mr,\mr')=\int_{0}^{1}c^{\rcoul}_{ij}(\mr,\mr';\lambda)d\lambda.
\label{cijeff}
\end{equation} 
Even though the relations (\ref{ci1}) and (\ref{cijeff}) are formally exact, they require knowledge of of the direct pair correlation functions for inhomogeneous systems with densities between reference and the equilibrium state, which in general are not known. 
%We notice that the introduction of interpolation parameter $\lambda$ that ``tunes'' the distribution functions to desired equilibrium profiles is similar in spirit to the traditional Debye charging process, in which the ionic charges continuously change until their final values are reached \cite{Le02}. Here, however, the physical mechanisms behind the changes in $\lambda$ are less obvious. In the framework of DFT, introduction of such charging parameter in the ionic interactions leads to an exact relation for the excess electrostatic functional in terms of the total pair correlation function.
Further progress can be achieved by removing the long-distance Coulomb interaction from the pair direct correlation functions in Eq. (\ref{cijeff}), which are expected to behave as $c_{ij}^{\rcoul}(\mr,\mr'; \lambda)\sim - \lambda_{B}\alpha_{i}\alpha_{j}/|\mr-\mr'|$ at large particle separations. The resulting short ranged direct correlation $c_{ij}^{\rsr}(\mr,\mr'; \lambda)= c_{ij}^{\rcoul}(\mr,\mr'; \lambda)+\lambda_{B}\alpha_{i}\alpha_{j}/|\mr-\mr'|$ can be used together with Eq. (\ref{cijeff}) to rewrite Eq. (\ref{ci1}) as:
\begin{equation}
\delta c_{i}^{\rcoul}(\mr)=-\alpha_{i}\delta\Phi(\mr)+\delta c_{i}^{\rres}(\mr),
\label{ci2}
\end{equation} 
where $\delta\Phi(\mr)$ is the electrostatic potential arising when the ionic density profiles in the reference state are perturbed by an amount $\delta\rho_{i}(\mr)$. In the second term on the right-hand side of Eq.(\ref{ci2}), we have defined the {\it residual} direct correlation function as, 
\begin{equation}
\delta c_{i}^{\rres}(\mr)=\sum_{j}\int\delta\rho_{j}(\mr')\bar{c}^{\rsr}_{ij}(\mr,\mr')d\mr',
\label{ci3}
\end{equation}
where the mean short range correlation $\bar{c}^{\rsr}_{ij}(\mr,\mr')$ is defined in Eq. (\ref{cijeff}), with the integral in $\lambda$ performed over $c_{ij}^{\rsr}(\mr,\mr';\lambda)$. 
%Eq. (\ref{ci2}) can be interpreted as the change in the work against the electrostatic interactions which is necessary in order to bring one ion of component $i$ at position $\mr$, as the ionic profiles change from the reference to the final equilibrium state. 
Clearly, the first term on the right-hand side of (\ref{ci2}) corresponds to the mean-field electrostatic potential, while the residual electrostatic single-particle correlations contain all the correlation effects beyond the mean-field.  
Combining the above results with the Euler-Lagrange equation, Eq. (\ref{dist2}), the ionic profiles corresponding to an arbitrary external potential $\phi_{i}(\mr)$ acting on the ionic system can be finally written as:
\begin{equation}
\rho_{i}(\mr)=\rho_{i}^{0}(\mr)\exp[-\beta\delta\phi_i(\mr)-\alpha_{i}\delta \Phi(\mr)+\delta c_{i}^{\rhc}(\mr)+\delta c_{i}^{\rres}(\mr)].
\label{dist3}
\end{equation}
In the case of planar double layers, where the external potential is provided by the electrostatic and hard core ion-wall interactions, the density profiles change only along the $z$ direction perpendicular to the parallel plates, and Eq. (\ref{dist3}) reduces to:    
\begin{equation}
\rho_{i}(z)=\rho_{i}^{0}(z)\exp[-\alpha_{i}\delta\psi(z)+\delta c_{i}^{\rhc}(z)+\delta c_{i}^{\rres}(z)],
\label{dist4}
\end{equation}
where now $\delta\psi(z)$ represents the (dimensionless) total electrostatic potential change with respect to the reference state (including the ionic interactions with the wall). Here, it is assumed that the hard core ion-wall interaction is the same for both equilibrium and the reference state. 
Eq. (\ref{dist4}) represents the basic relation for the planar electric double layer structure in the framework of the DFT approach. In practice, it has to be coupled with the Poisson equation for the ionic distributions. In the case depicted in Fig.~\ref{fig:fig1} of two parallel charged plates located at positions $\pm d/2$, Poisson equation is:
\begin{equation}
\dfrac{d^2\psi}{dz^2}=-4\pi\lambda_{B}\biggr[\sum_{i}\rho_{i}(z)+\sigma[\delta(z+d/2)+\delta(z-d/2)]\biggr],
\label{poisson}
\end{equation}
where $\sigma\equiv\sigma_s/q$ is the surface density of charge carriers on the wall. Once the single-particle hard-core and residual electrostatic correlations are known, Eqs. (\ref{dist4}) and (\ref{poisson}) can be self-consistently solved (e. g. via a Picard iteration method) to provide the ionic distributions around the charged plates. As already mentioned, the exclusion volume contributions can be evaluated with a high degree of accuracy in the framework of the FMT approach. What essentially determines the ability of different DFT approaches to capturing the fine structure of the EDL are, therefore, the approximations employed in the evaluation of both reference state and the residual electrostatic correlations. Several different approaches have been proposed over the last years to accurately calculate these quantities in a number of different contexts \cite{Yan15}. Since very little is actually known about the pair correlations of inhomogeneous ionic system, it is in practice  usual to take a homogeneous system, with the reservoir bulk concentrations $\bar{\rho_{i}}$, as a representative reference state. In most cases, a further bulk-like approximation is invoked by taking the effective pair correlation functions in Eq. (\ref{ci3}) to coincide with the corresponding bulk ones:
\begin{equation}
\bar{c}_{ij}^{\rsr}(|\mr-\mr'|)\approx c_{ij}^{\rsr}(|\mr-\mr'|;\{\bar{\rho}_{i}\}),  
\label{bulk}
\end{equation}
where the correlations on the right-hand side are calculated in the charge-neutral bulk fluid. This approximation, usually referred to in the literature as the {\it bulk expansion approximation}, can be easily identified as a second-order truncation in the Taylor functional expansion of the excess coulomb free energy around the reference bulk state \cite{Jai14}. It was first employed by Rosenfeld \cite{Ros93} more than twenty years ago, and has been extensively used since then to describe EDL structures. The bulk expansion approximation is also equivalent to the {\it hyppernetted chain} (HNC) approximation for the residual wall-ion correlation \cite{Hansen}. It is employed in a combination with the Mean Spherical Approximation (MSA) for the bulk direct pair correlations (bulk-MSA) for ion-ion interactions, for which an analytical solution is known exactly \cite{Blum81}. This leading-order bulk-MSA approximation has been applied with a good success in a number of situations involving both planar \cite{Pat10,Hen011,Wu11,
Med14,Zho14} as well as spherical \cite{Yu04,Li04} double layers, provided the electrostatic coupling inside the EDL is not to high \cite{Jai14,Yan15}. 

Going beyond the bulk-MSA approach, Gillespie {\it et al.} \cite{Gil02,Gil03,Gil05} proposed a Reference Density Fluid (RDF) approximation in which the residual electrostatic correlations are expanded around an inhomogeneous reference fluid represented by density profiles that are themselves functionals of the real equilibrium profiles. Having recognized that these profiles do not need to represent real equilibrium states of the system, they have chosen them in such a way as to satisfy electroneutrality at each position, allowing for the use of a local approximation for the evaluation of the pair correlation functions, again in the context of the MSA approach. Another accurate approach for the residual contributions is the so-called Weighted Correlation Approach (WCA), which has been recently developed by Wang {\it et al.} \cite{Wan11,Wan11_2,Jia14}. The basic idea relies on the interpretation of the effective direct correlations in Eq. (\ref{cijeff}) as the correlations which are averaged over the path of 
inhomogeneous densities connecting the reference and the final state. Since the residual correlations are short ranged, the resulting mean correlations in Eq. (\ref{ci3}) can be approximated by direct pair correlations {\it weighted} over a region whose typical size is the range of the residual correlations \cite{Wan11}. The question of which approximation -- among bulk-MSA, RDF and WCA -- is best to describe the EDL structure seems to depend on the particular problem at hand \cite{Jai14}, although the bulk-MSA is in all cases computationally cheaper than the other ones. Recently, Yang and Lui performed a careful analysis on the accuracy of these approaches within a large range of electrostatic couplings. They conclude that the optimal choice is in general a combination of these methods \cite{Yan15}, in which the reference profiles used in Eq. (\ref{dist4}) and the ones used for calculating $\delta c_{ij}^{\rres}(\mr)$ in Eq. (\ref{ci3}) are decoupled from one another, leading to different approximations 
for the zeroth and first order residual correlations \cite{Yan15}. Besides all these expansion-based approaches, in a very recent contribution Roth and Gillespie \cite{Roth16_2} proposed a first-principle approach in which the residual single-particle correlations are obtained via a generalization of the MSA bulk excess free energy to an inhomogeneous system, showing promising results with yet relatively low computational cost. 

All the aforementioned approximations for the residual contributions involve the application of the MSA in the calculation of the direct pair ion-ion correlations. Since the MSA is based on a linear response approximation for such correlations, it is reasonable to expect that such MSA-based approaches should fail at sufficiently high electrostatic couplings -- where the MSA proves to be inaccurate even for bulk solutions \cite{Lev96}. In these limits, strong non-linear effects such as ionic association start to take place \cite{Le02,Pai11}, therefore requiring more sophisticated approaches for the ionic correlations. One easy way to circumvent this problem is to adopt an approximation in which the electrostatic bulk correlations are evaluated in the framework of the HNC relation, whereby the total and direct pair correlations are related by
\begin{equation}
h_{ij}(r)=\exp[-\beta u_{ij}(r)+h_{ij}(r)-c_{ij}(r)]-1,
\label{hnc}
\end{equation}
where $h_{ij}(r)$ is the radial total correlation function. The MSA relation $c_{ij}(r)=-\beta u_{ij}(r)$ is clearly recovered upon linearization of the exponential factor on the right-hand side. This relation is complemented by the Ornstein-Zernike (OZ) equation for the homogeneous bulk electrolyte, which in the Fourier space reads as:
\begin{equation}
\hat{h}_{ij}(k)=\hat{c}_{ij}(k)+\sum_{l}\hat{c}_{il}\bar{\rho}_{l}\hat{h}_{lj}(k).
\label{OZ}
\end{equation} 
These relations can be numerically evaluated to obtain the ionic direct correlation functions $c_{ij}(r)$ for the bulk system. Although remarkably more accurate than its MSA counterpart, the HNC approach has the disadvantage of not possessing an analytic solution, which makes its direct application in the context of RDF or WCA approaches quite difficult from a computational perspective. However, the implementation of the HNC approximation for the electrostatic correlations together with bulk expansion, Eq. (\ref{bulk}), can be readily accomplish with not significant additional increase in computational effort. Since we aim to investigate the EDL at high electrostatic couplings -- namely high ionic valencies and small surface charges -- we will, in what fallows, employ (unless otherwise specified) the bulk expansion in combination with the HNC solution obtained from Eqs. (\ref{hnc}) and (\ref{OZ}) for computing  
the residual contributions in Eq. (\ref{ci3}). While for moderate ionic correlations the  MSA-based approach should provide results comparable  to the HNC-based approximation, we expect HNC to be much more accurate in the limit of very strong electrostatic correlations -- where the validity of even bulk MSA approach must to be put in doubt.

\section{Results}

Having established the theoretical basis as well as the simulation methods to be employed, we now briefly outline some aspects of their numerical implementation before proceeding to analyze the properties of the model electrolyte system described in Sec. II. 

In practice, the regions outside the charged plates --- the bulk --- are taken to be large enough to guarantee that the ionic profiles relax to their bulk values sufficiently far away from the plates. Using these bulk concentrations as the reference state, the ionic profiles in Eq. (\ref{dist4}) read as:  
\begin{equation}
\rho_{i}(z)=\bar{\rho}_{i}\exp(-\alpha_i\delta\psi(z)+\phi_{i}^{\rhc}(z)+\delta c_{i}^{\rhc}(z)+\delta c_{i}^{\rres}(z)).
\label{rho5}
\end{equation}
The bulk densities $\bar{\rho}_{i}$ are set by the reservoir salt concentration, and should satisfy the overall electroneutrality condition $\sum_{i}\rho_{i}\alpha_i=0$. The potential $\phi_{i}^{\rhc}(z)$ represents the exclusion volume interaction between the ions and the hard walls located at positions $z=\pm d/2$:
\begin{eqnarray}
\phi_{i}^{\rhc}(z) & = & \begin{cases} \infty,\qquad   -a_{i} \leq (|z|-L/2) \leq a_{i},
\\ 0,\qquad\qquad {\rm otherwise}.
\end{cases}
\label{phi_hc}
\end{eqnarray}
Making use of the one-dimensional Poisson equation, Eq. (\ref{poisson}), the difference between inhomogeneous and bulk potentials $\delta\psi(z)=\psi(z)-\psi_b$ can be conveniently written as a simple functional of the ionic profiles as:
\begin{equation}
\delta\psi(z)=\beta q\phi^{\rcoul}_p(z)+4\pi\lambda_B\sum_{i} \alpha_{i}\int_{-z_b}^{z}(z'-z)\rho_{i}(z')dz',
\label{poisson_int}
\end{equation}
where $\phi^{\rcoul}_p(z)$ is the electrostatic potential produced by the charged walls, as defined in Eq.(\ref{elec}), and $z_{b}$ represents a position deep inside the bulk solution, $|z_{b}|\gg d/2$. When the single-particle correlations are set to zero in Eq. (\ref{rho5}), combination with Eq. (\ref{poisson_int}) recovers the traditional Poisson-Boltzmann equation for the ionic distributions. Since both residual and hard sphere contributions are also explicitly written as functionals of the ionic profiles, Eqs. (\ref{rho5}) and (\ref{poisson_int}) can be numerically solved with a standard Picard-like iteration method. Starting with guess functions $\rho_{i0}(z)$, the corresponding electrostatic profiles are calculated from Eq. (\ref{poisson_int}), along with the FMT hard sphere contributions $\delta c_{i}^{\rhc}(z)$, as described in Appendix A. As for the residual single-particle correlations $\delta c_{i}^{\rres}(z)$, the bulk fluid expansion resulting from the application of Eqs. (\ref{bulk}) and (\ref{ci3}) is employed. Here, the direct pair correlations for the bulk system are calculated via the solution of the OZ equation, Eq. (\ref{OZ}), together with either MSA or the HNC approximations, Eq. (\ref{hnc}). While for the MSA case analytic expressions for the pair correlations are available, the HNC approach requires numerical solutions, which are here obtained using the methods described in Ref. [\onlinecite{Ng74}]. For the computation of $\delta c_{i}^{\rres}(z)$, one has to explicitly remove the short range hard sphere contributions from the HNC solution. This is accomplished by taking $c_{ij}^{\rcoul}(r)=c_{ij}(r;\lambda_B)-c_{ij}(r;\lambda_B=0)$ for the HNC electrostatic correlations. Once all the relevant contributions are calculated, new estimate density profiles are obtained using Eq. (\ref{rho5}). The whole process is then repeated until convergence is achieved. In practice, improved estimates for the ionic profiles are obtained by properly combining the corresponding input and output resulting from 
consecutive iterations. Finally, we notice that the imposed boundary conditions -- namely that the bulk concentrations should be recovered far away from the charged walls, $\rho_{i}(|z|\rightarrow z_{b})\rightarrow \bar{\rho_{i}}$, are naturally satisfied within this iterating scheme, provided they are fulfilled by the initial profile guess.

We are now going to use the DFT theory presented above to 
study ionic distributions between two like-charged surfaces in a contact with a bulk salt reservoir, focusing on the equilibrium ionic profiles, degree of charge neutrality breakdown between the charged walls, 
and the net force on each surface.

\subsection{Ionic profiles}

\begin{figure}[h!]
\centering
\includegraphics[width=6.5cm,height=4.5cm]{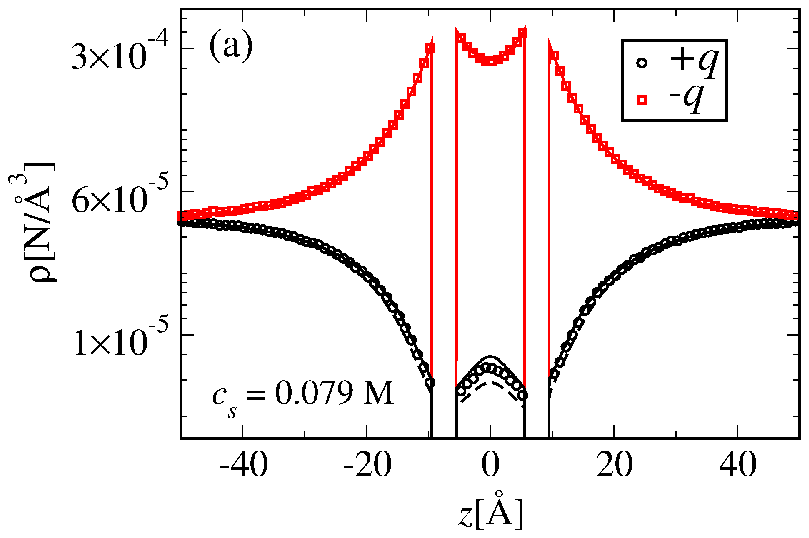}
\includegraphics[width=6.5cm,height=4.5cm]{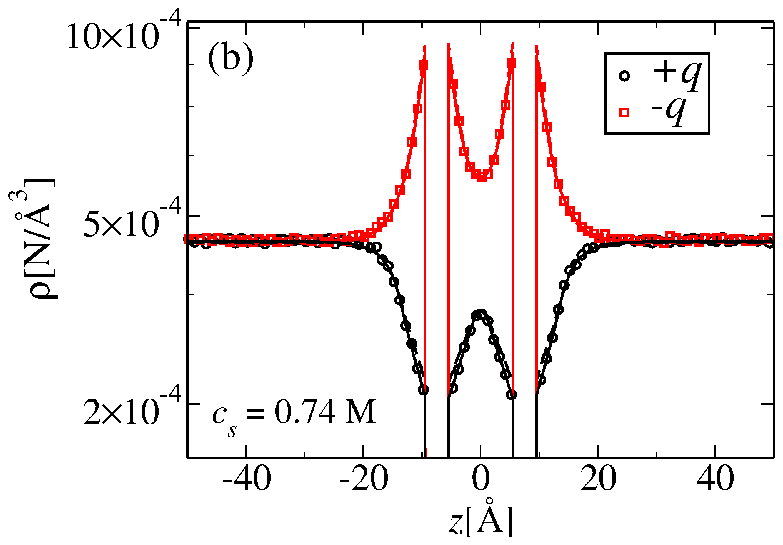}\\\vspace{1cm} 
\includegraphics[width=6.5cm,height=4.5cm]{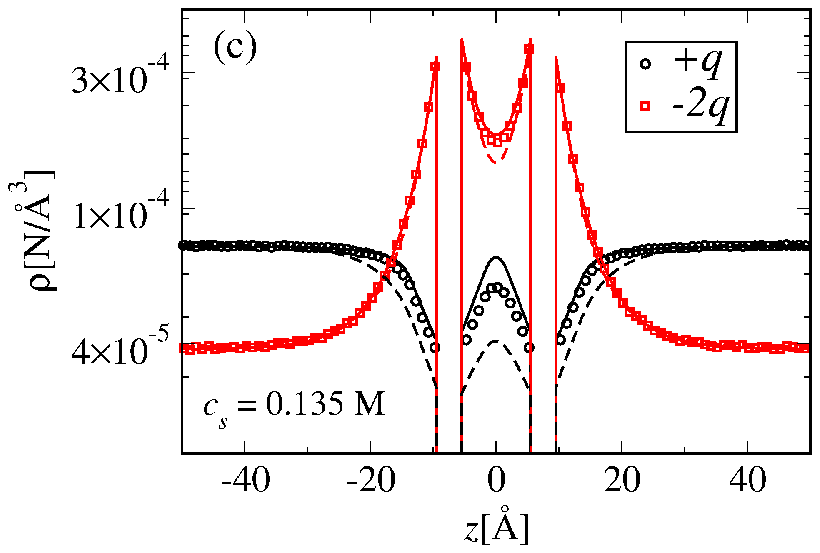} 
\includegraphics[width=6.5cm,height=4.5cm]{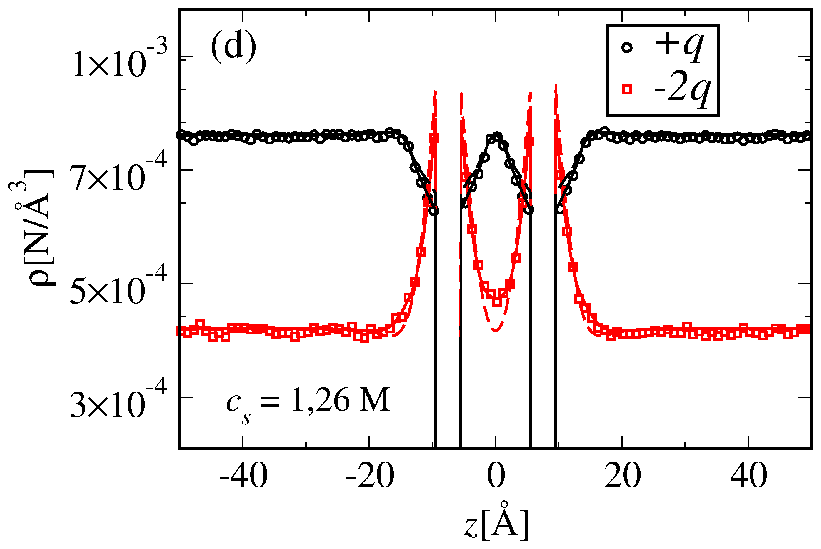}\\\vspace{1cm}
\includegraphics[width=6.5cm,height=4.5cm]{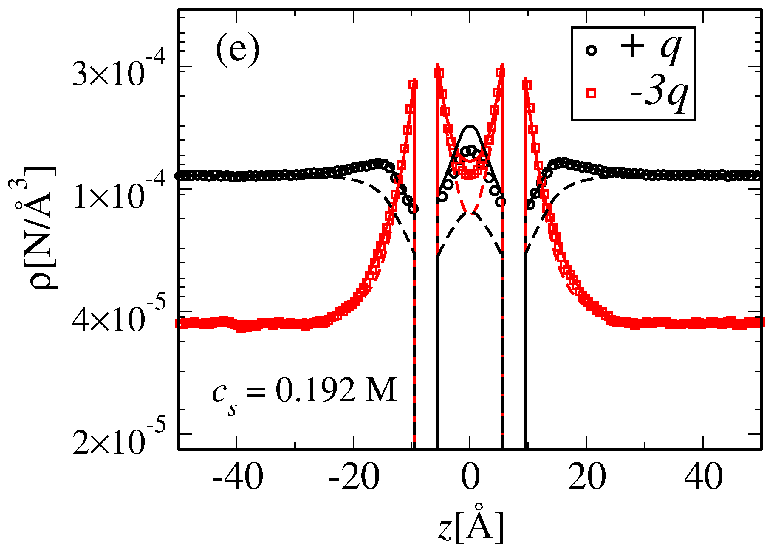}
\includegraphics[width=6.5cm,height=4.5cm]{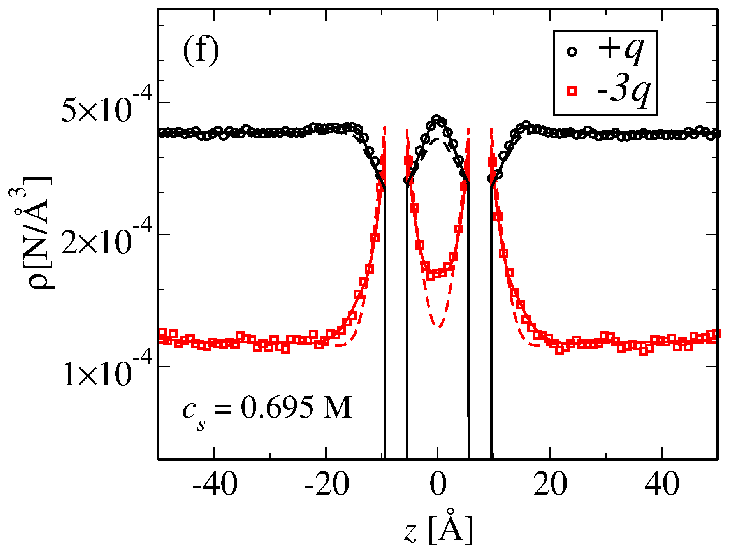}
\caption{Equilibrium ionic profiles for various counterion valences and salt concentrations, for charged walls at a separation $d=15$~\AA~ from one another. Open circles and squares represent the MC cationic and anionic profiles, respectively. The dashed lines are theoretical results using the bulk-MSA approximation, while the full curves correspond to profiles obtained within the bulk-HNC approach. In all cases, the particle radii are fixed at $a_{+}=a_{-}=2$~\AA, while the electrolyte charge asymmetries are 1:1 (a) and (b), 2:1 (c) and (d), and 3:1 (e) and (f). The corresponding reservoir salt concentrations are displayed in the figures. From (a) to (d), the wall surface charge density is fixed at $\sigma_s=0.064~\mathrm{C/m^2}$, while in (e) and (f) it is $\sigma_s=0.0576~\mathrm{C/m^2}$.}
\label{fig:fig2}
\end{figure}

Ionic distributions around two charged surfaces separated by distance $d=15$ \AA~ are shown in Fig.~\ref{fig:fig2}, for several salt concentrations and electrolyte asymmetries. Open symbols represent the MC results, the solid lines are profiles obtained from the bulk-HNC approximation, while dashed lines stand for the bulk-MSA approach. We note that the recently developed efficient MC method \cite{San16}, allows to very accurately perform statistics while using small grid sizes over the simulation box. The resulting ionic profiles are able to precisely capture the fine details of the double layer structure in the vicinity of the charged surfaces, providing an excellent basis for testing the validity of different theoretical approaches.

%%%%%%%%%%%%%%%%%%%%%%%%%%% Fig. 2 %%%%%%%%%%%%%%%%%%%%%%%%%%%%%%%%%      
%%%%%%%%%%%%%%%%%%%%%%%%%%%%%%%%%%%%%%%%%%%%%%%%%%%%%%%%%%%%%%%%%%%%%

Ionic density profiles are strongly inhomogeneous in the region between the plates, and assume their bulk values shortly beyond the charged walls. Exception is the 1:1 electrolyte at small salt concentration (Fig.~\ref{fig:fig2}a), in which case  the ionic inhomogeneities extends farther away from the walls, which is clearly a consequence of a larger screening length. We can also see that both bulk-HNC and bulk-MSA show excellent agreement with the simulations in the case of monovalent electrolyte (Fig.~\ref{fig:fig2}a and Fig.~\ref{fig:fig2}b). Since the curves overlap each other, it is actually difficult to distinguish between the two approximations. At such weak electrostatic couplings, the hard core correlations strongly dominate over the electrostatic ones, and the ionic density profiles are well described by the mean-field electrostatic approximation in addition to hard-core corrections \cite{Fry12}. As the electrostatic coupling increases -- changing the anionic valencies (2:1 in Fig.~\ref{fig:fig2}c 
and Fig.~\ref{fig:fig2}d and 3:1 in Fig.~\ref{fig:fig2}e and Fig.~\ref{fig:fig2}f) -- the deviations between different theoretical approaches begin to appear, the bulk-HNC being always the closest  to the simulation results. In parti\-cular, at higher salt concentrations and strong electrostatic coupling (Fig.~\ref{fig:fig2}d and Fig.~\ref{fig:fig2}f) the bulk-MSA results fail to reproduce the MC counterion and coion distributions, while the bulk-HNC approximation shows perfect agreement with the simulations. The situation is slightly different at smaller salt concentrations (see Fig.~\ref{fig:fig2}c and Fig.~\ref{fig:fig2}e), where still the bulk-MSA deviates from the MC data at strong couplings. Although closer to the MC results, the bulk-HNC approximation in this situation is less accurate in comparison with the high salt case. This loss of accuracy is a consequence of the local character of the bulk expansion. While the HNC approach can correctly account for the stronger magnitude of ionic correlations 
in the case of divalent and trivalent ions, at small ionic strengths the range of the electrostatic correlations becomes larger and, therefore, more important, rendering the bulk expansion less accurate. It is likely that in this limit the non-local approaches such as the RDF or the weighted densities approximations -- in which the inhomogeneities are averaged over a distance that scales with the range of the electrostatic correlations -- will become more reliable for describing electrostatic correlations. Unfortunately, the implementation of such methods in a combination with the HNC 
is very difficult. A detailed comparison of the different DFT approaches against the MC results at several salt concentration and ionic asymmetries goes beyond the scope of this work, and will be the subject of future investigations. For now, we simply emphasize that the bulk-HNC is far more accurate than the bulk-MSA approximation and, therefore, in what follows we will exclusively use the bulk-HNC approach in the calculation of the system properties.

\subsection{Electroneutrality}

It is not clear from the density profiles in Fig.~\ref{fig:fig2} whether or not the electrolyte confined between the charged plates obeys charge neutrality. Notice that nowhere this condition has been explicitly imposed in our calculations. We now investigate under what circumstances will the electroneutrality breakdown takes place. To this end, we define the total charge of the confined electrolyte per unit area as:
\begin{equation}
\sigma_{\rin}=\int_{-d/2}^{d/2}[\alpha_{+}\rho_{+}(z)-\alpha_{-}\rho_{-}(z)]dz=2\int_{0}^{d/2}[\alpha_{+}\rho_{+}(z)-\alpha_{-}\rho_{-}(z)]dz.
\label{sigma_in}
\end{equation}

%%%%%%%%%%%%%%%%%%%%%%%%%%% Fig. 3 %%%%%%%%%%%%%%%%%%%%%%%%%%%%%%%%%
\begin{figure}
\centering
\vspace{0.5cm}
\includegraphics[width=5.4cm,height=4.25cm]{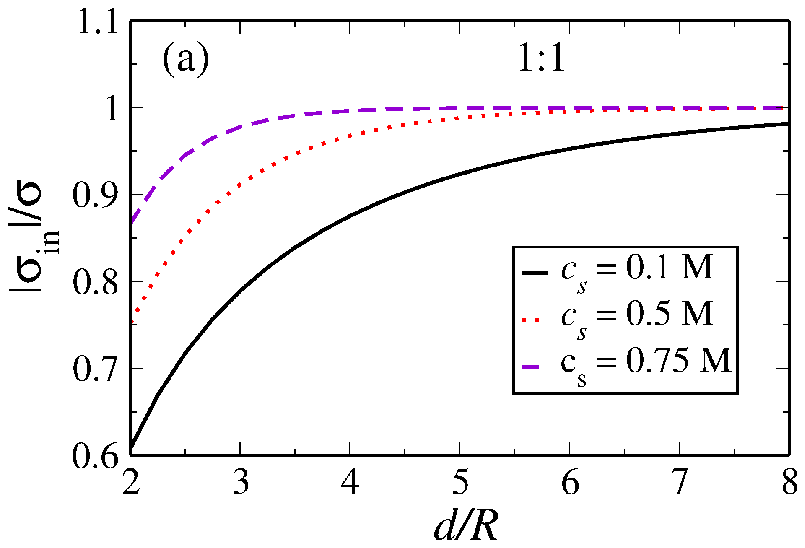}
\includegraphics[width=5.4cm,height=4.25cm]{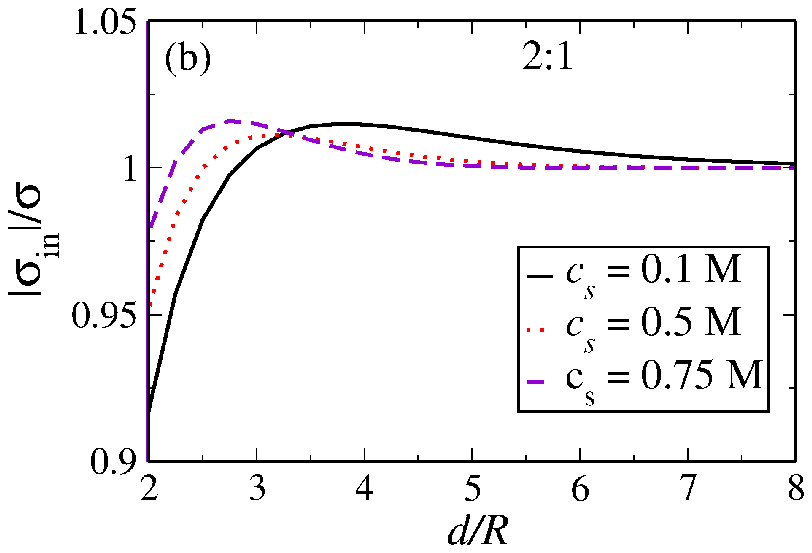} 
\includegraphics[width=5.4cm,height=4.25cm]{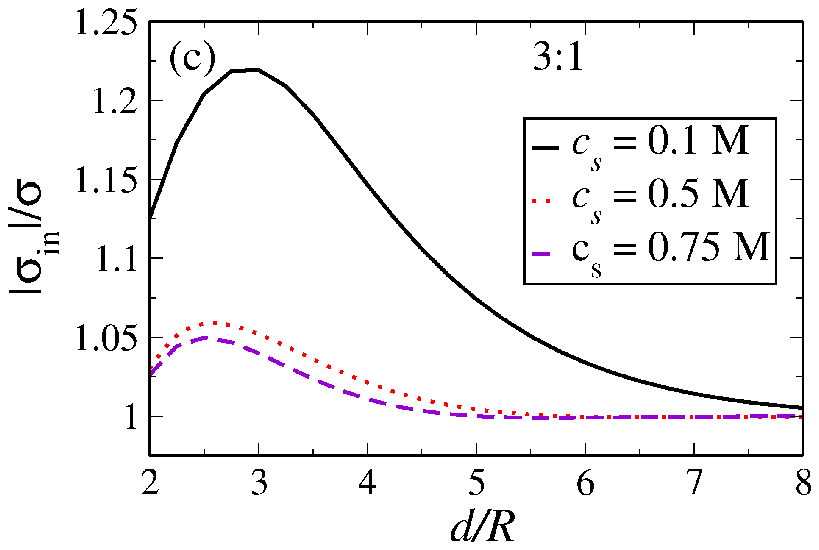}
\caption{Ratio between the magnitudes of net charge per unit area confined between the surfaces, $\sigma_{\rin}$, and the surface charge density, $\sigma$, as a function of the plate's separation for different electrolyte asymmetries and salt concentrations. In all the cases, the surface charge density on the plates is $\sigma_s=q\sigma=0.0576~\mathrm{C/m^2}$, and the ionic radii are $a_{+}=a_{-}=2$ \AA. The wall separation $d$ is scaled in terms of the effective diameter $R \equiv a_{+}+a_{-}$.}
\label{fig:fig3}
\end{figure}
%%%%%%%%%%%%%%%%%%%%%%%%%%%%%%%%%%%%%%%%%%%%%%%%%%%%%%%%%%%%%%%%%%%%%

The, electroneutrality between the plates, $-d/2<z<d/2$ is satisfied if $Q_t=\sigma+\sigma_{\rin}=0$, see Fig.~\ref{fig:fig1}. In Fig.~\ref{fig:fig3}, the ratio between the magnitude of such ``internal'' net surface charge and the bare plate surface charge is plotted as a function of the wall separation for different electrolyte asymmetries and salt concentrations. In all cases, breakdown of charge neutrality is found at small wall separations. As the distance between the walls grows larger, the ratio $|\sigma_{\rin}|/\sigma$ converges rapidly to unity, meaning that the overall electroneutrality in the region between the charged walls is naturally recovered. The crossover distance at which charge neutrality starts to take place is dependent on the amount of salt in the bulk reservoir. Clearly, at small salt concentrations (full curves) larger wall separations are required to guarantee charge neutrality in the interior region. Again, this can be understood in terms of the larger inhomogeneity region across 
the charged walls at smaller ionic strengths, as has been observed in the left panels of Fig.~\ref{fig:fig2}.  As the salt concentration increases, the ionic profiles rapidly converge to the bulk value in the outer region, leading to a decrease in neutrality breakdown between the plates. This effect is also consistent with the Donnan approach in which the electrostatic potential difference between the system and the reservoir 
becomes smaller as the salt concentration increases \cite{Tam98,Ohshi85}.

There are several physical mechanisms responsible for electroneutrality violation in narrow pores. First, electrostatic effects lead to strong counterion condensation at the charged walls, in an attempt to neutralize their surface charge. On the other hand, entropic effects try to induce a homogeneous particle distribution all over the system, thereby limiting counterion condensation. Electrostatic correlations between anions and coions are responsible for complex ionic association, which also effectively reduce the amount of counterion condensation. Moreover, the bulk counterion-counterion correlations favor additional condensation.  Finally, exclusion volume correlations strongly constrain the number of ions which can be present between the walls, significantly affecting the charge neutrality of the interior region. Clearly, this effect is much more important for small wall separations, when only 
few layers of ions can be accommodated  between the surfaces. 

Fig.~\ref{fig:fig3} also reveals the effect of ionic correlations on electroneutrality. As the counterion valency changes, different qualitative behaviors are clearly observed. In the case of monovalent ions (Fig.~\ref{fig:fig3}a), the overall charge between the surfaces grows monotonically as the wall separation increases. Strong violation of charge neutrality is found in this regime, the net charge between the walls being always positive ($|\sigma_{\rin}|\le\sigma$). Remarkably, at the smallest salt concentration $c_{s}=0.1$ M and at the narrowest separation $d/R=2$ studied, the net internal charge density is $40\%$ of the bare wall surface charge. Upon salt addition, the magnitude of electroneutrality breakdown becomes weaker, and the separation between the surfaces at which it occurs shorter. A quite different behavior takes place for higher charge asymmetries, as can be observed in Fig.~\ref{fig:fig3}b and Fig.~\ref{fig:fig3}c. For both 2:1 and 3:1 electrolyte, the magnitude of the net charge 
between the plates reaches a maximum before charge neutrality is achieved. Furthermore, the phenomenon of charge reversal -- whereby the internal double layer has a net charge with sign opposite to the surface charge -- is observed in these cases and is a manifestation of strong electrostatic correlations between the multivalent counterions.   In the case of 2:1 electrolyte (Fig.~\ref{fig:fig3}b)  for very narrow slits the internal charge density is smaller than the surface charge.  However, when the separation between the surfaces is increased 
the internal charge overcompensates the surface charge, resulting in a charge reversal.  For larger separation between the plates the charge neutrality is restored.  The situation is remarkably different for trivalent counterions (Fig.~\ref{fig:fig3}b), for which the internal region is always overcharged for very narrow pores -- the internal net charge has a sign opposite to the  bare surface charge ($|\sigma_{\rin}|\ge\sigma$) at salt concentrations $c_{s}=0.1$ M and $c_{s}=0.5$ M. On the other hand, when $c_{s}=0.75$ M (dashed curve) the internal region is undercharged for intermediate separations  -- the net EDL charge is slightly positive when $5\le d/R\le 7$. Another interesting feature in the case of trivalent counterions is that the breakdown of charge neutrality is significantly more pronounced for very low salt concentration, $c_{s}=0.1$ M, and narrow slits. 

%%%%%%%%%%%%%%%%%%%%%%%%%%% Fig. 4 %%%%%%%%%%%%%%%%%%%%%%%%%%%%%%%%%
\begin{figure}
\centering
\vspace{0.5cm}
\includegraphics[width=5.4cm,height=4.25cm]{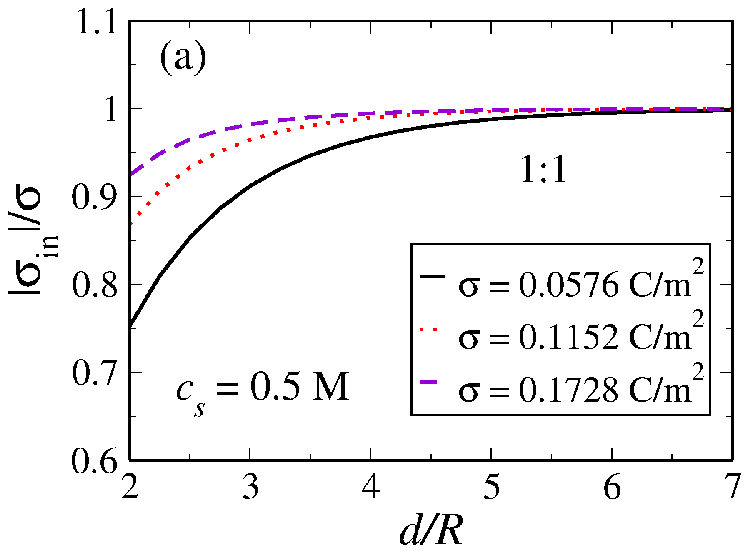}
\includegraphics[width=5.4cm,height=4.25cm]{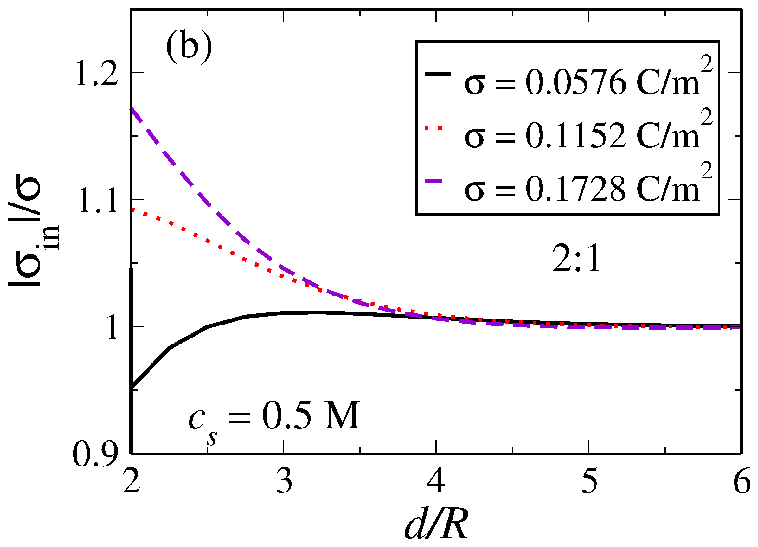} 
\includegraphics[width=5.4cm,height=4.25cm]{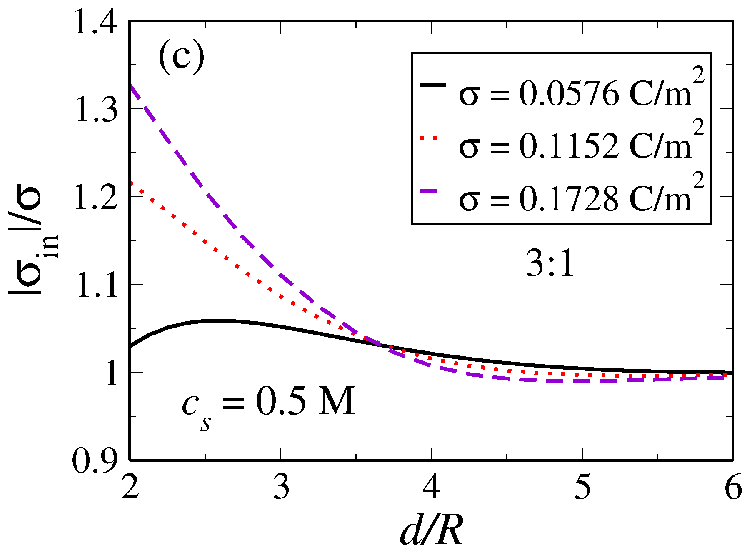}
\caption{Ratio between the magnitudes of  charge per unit area of the confined electrolyte, $\sigma_{\rin}$, and the surface charge $\sigma$, as a function of the plate separation for different bare surface charges $\sigma$ and electrolyte  asymmetries 1:1 (a), 2:1 (b) and 3:1 (c). In all the cases, the reservoir salt concentration is $c_s=0.5$ M, and the ionic radii are $a_{+}=a_{-}=2$~\AA. Once again, the wall distances $d$ are scaled with the effective diameter $R\equiv a_{+}+a_{-}$.}
\label{fig:fig4}
\end{figure}
%%%%%%%%%%%%%%%%%%%%%%%%%%%%%%%%%%%%%%%%%%%%%%%%%%%%%%%%%%%%%%%%%%%%%

To investigate the effect of increasing the surface charge at constant salt concentrations on 
charge neutrality violation, the ratio $|\sigma_{\rin}|/\sigma$ as a function of the wall separation for the different charge asymmetries is plotted in Fig.~\ref{fig:fig4}. Once again, a different qualitative behavior is observed in the case of monovalent and multivalent ions (Fig.~\ref{fig:fig4}a). While a system with 1:1 electrolyte always remains undercharged, and restores
the internal charge neutrality only at asymptotically large separations, both 2:1 and 3:1 electrolytes exhibit strong charge reversal for very narrow pores (Fig.~\ref{fig:fig4}b and Fig.~\ref{fig:fig4}c). For monovalent electrolytes the degree of electroneutrality breakdown becomes weaker as the  surface charge increases. On the other hand for very narrow slits, we see that while a system with 2:1 electrolyte is undercharged, for very low surface charge densities, it actually becomes overcharged for large charge density.  For 3:1 electrolyte, the overcharging occurs even at very low surface charge densities, and becomes very pronounced as the surface charge is increased.

%Although seemingly counter-intuitive, we should notice that this result does not imply that the {\it magnitude} of the net charge is decreasing when the bare charge becomes larger. Since the bare charges increases by a factor of two in the different curves, the total charge confined between the walls, $\sigma(1-|\sigma_{\rin}|/\sigma)$, actually {\it grows} as the bare charge increases. This is however more pronounced in the case of multivalent counterions. Here another different behavior is also evident: the ratio $|\sigma_{\rin}|/\sigma$ no longer achieves a maximum at larger surface charges in the presence of multivalent ions (as observed in Fig.~\ref{fig:fig3}). Instead, it decreases monotonically for the $2:1$ electrolyte, and reaches a minimum at large separations in the case of $3:1$ electrolyte. 
%It is also interesting to note that the net charge assumes always a sign opposite to that of the charged plate in the case of $2:1$ electrolytes at large surface charges (see dotted and dashed curves in Fig.~\ref{fig:fig4}b).

\subsection{Wall forces}

We now investigate the interaction between the two surfaces. According to the Contact Value Theorem (CVT), the net force acting on each wall has two contributions, an electrostatic 
and an entropic.  The electrostatic force felt by one of the surfaces 
is produced by the interaction with the other surface and with the electrolyte, both external and internal ones.  The entropic force arises from direct collisions of ions with the surface and the resulting momentum transfer.  

Consider the surface located at $-d/2$. The electric field it produces is $E(z)=2\pi\lambda_{B}\sigma\text{sign}(z+d/2)$, (where $\text{sign}(x)$ denotes the sign of $x$).  Newton's third law then requires that 
the net force per unit of area, $F/A$, is
\begin{equation}         
\beta \Pi\equiv\beta F/A=2\pi\lambda_B\sigma(\sigma_{>}-\sigma_{<})+\sum_{i}\rho_{i}(-d/2+a_i)-\rho_{i}(-d/2-a_i),
\label{F1}
\end{equation}
where $\sigma_{>}$ and $\sigma_{<}$ represent the total charge per unit of area located in the regions $z>-d/2$ and $z<-d/2$, respectively. Since for equally charged surfaces the ionic distributions are even functions, $\rho_{i}(-z)=\rho_{i}(z)$, it is easy to see that these quantities are constrained by the condition $\sigma_{>}=\sigma_{\rin}+\sigma+\sigma_{<}$. The total force on the plate located at $z=-d/2$ can therefore be rewritten as:
\begin{equation}         
\beta \Pi\equiv\beta F/A=2\pi\lambda_B\sigma(\sigma+\sigma_{\rin})+\sum_{i}\rho_{i}(-d/2+a_i)-\rho_{i}(-d/2-a_i).
\label{F2}
\end{equation}
Notice that we have defined $\Pi$ in such a way that a positive force corresponds to repulsion between the charged surfaces. While the first term on the right-hand side of this equality represents the net electrostatic force per unit of area, the second term is the pressure resulting from the transfer of momentum from the ions to the surface. We notice here that in the present situation the same result is obtained from a direct thermodynamic calculation of the osmotic pressure across the walls, and that it does not depend on the particular approximations employed in building up the excess functional (see Appendix B). 
%Remarkably, the contact CVT applied at the charged walls will be therefore always thermodynamically consistent, no matter what approximations have been employed for the free energy functional. 

\begin{figure}
\centering
\vspace{0.5cm}
\includegraphics[width=5.4cm,height=4.25cm]{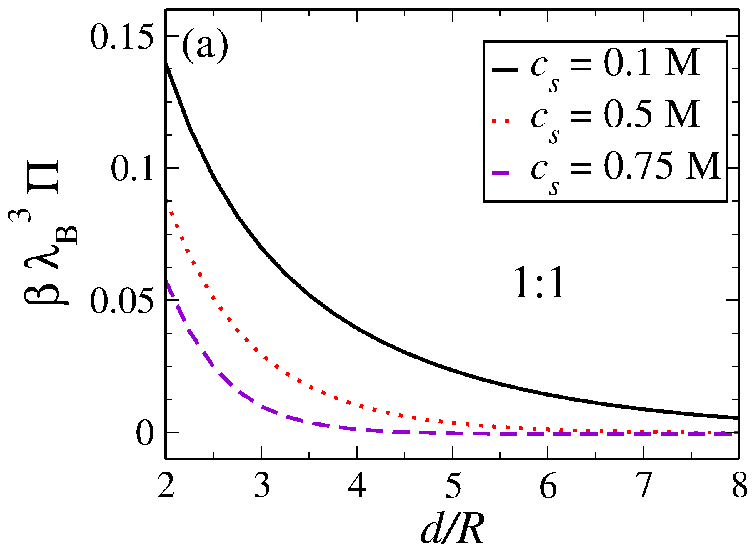}
\includegraphics[width=5.4cm,height=4.25cm]{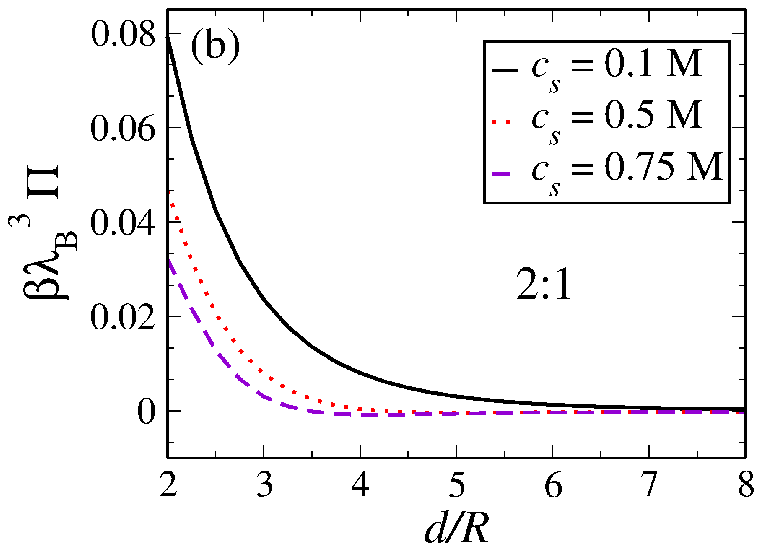} 
\includegraphics[width=5.4cm,height=4.25cm]{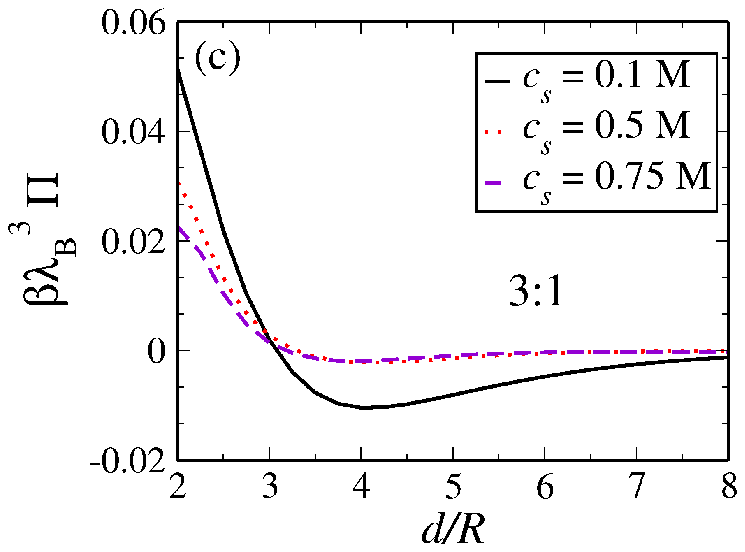}
\caption{The pressure $\beta \lambda_{B}^3\Pi$ on a charged walls for separation between the surfaces $d$. 
The system parameters are the same as in Fig.~\ref{fig:fig3}.}
\label{fig:fig5} 
\end{figure}

The net force, given by Eq. (\ref{F2}), shows a very non-trivial influence of the electroneutrality on the interaction between the surfaces. When $|\sigma_{\rin}|/\sigma<1$, the net charge between the walls is positive, leading to a repulsive electrostatic force. This, however, implies a net negative charge on the bulk side of the walls, which leads to a larger accumulation of counterions at the surfaces facing the bulk electrolyte. This imbalance of ions on both sides of the surface  
provides a net force that pushes the plates towards each other, leading to an attractive entropic contribution. In the opposite case where $|\sigma_{\rin}|/\sigma>1$, the inverse situation takes place: the electrostatic interactions become attractive, while the balance of ions between confined and bulk electrolytes leads to a repulsive force. It is the fine balance between these contributions -- described by the two terms at the right-hand side of Eq. \ref{F2} -- that dictates the net, ion mediated, force between the charged surfaces.
Fig.~\ref{fig:fig5} shows the net force per unit of area $\beta \lambda_{B}^3\Pi$ acting on a charged wall at different electrolyte charge asymmetries and salt concentrations. The parameters are the same as in Fig.~\ref{fig:fig3}. Although in this case the degree of electroneutrality for monovalent and multivalent ions has quite different behaviors, the net force between the walls has similar form for monovalent and divalent ions (Fig.~\ref{fig:fig5}a and Fig.~\ref{fig:fig5}b). In both cases, the force is always repulsive and decays monotonically when $c_{s}=0.1$ M and $c_{s}=0.5$ M. At the highest salt concentration $c_{s}=0.75$ M, the force becomes slightly attractive at wall separations $3\le d/R \le 5$ in the case of 2:1 electrolyte. In this region, the net charge within the walls is slightly positive, as can be verified in Fig.~\ref{fig:fig3}b which shows that $|\sigma_{\rin}|\lesssim \sigma$. The attractive force induced by the net momentum transfer by the bulk side ions, in this case, overcomes the weakly 
repulsive electrostatic force. The fact that such effect takes place at higher salt concentrations is due to stronger screening of the electric field on the bulk side of solution, which forces the ions to be accumulated within a narrow region in the vicinity of the wall surfaces (see Fig.~\ref{fig:fig2}). The absence of attractive force in the case of
surfaces surrounded by monovalent electrolyte has been proven in the framework of the Poisson-
Boltzmann theory \cite{Tri99,Tri00}. 
Since  for 1:1 electrolyte the electrostatic correlations play only a minor role, it is not
surprising that no attraction is found between the charged plates under such conditions. It is important to stress, however, that a steric-driven attractive force as well as charge reversal have been already reported in the case of size asymmetric,
strongly confined, monovalent electrolytes \cite{Gue11,Gue11_2}. In the case of trivalent counterions (Fig.~\ref{fig:fig5}c), the force changes sign at all concentrations under consideration. Again, the low salt situation $c_s=0.1$ M differs qualitatively from the other ones. In this case, the force achieves a well defined minimum at $d/R=4$, and is attractive even at large wall separations. This result is consistent with the large ratio $|\sigma_{\rin}|/\sigma$ observed in Fig.~\ref{fig:fig3}c, which implies a net negative charge and an electrostatic attraction that strongly overcomes the repulsive pressure provided by the confined electrolyte. 

%%%%%%%%%%%%%%%%%%%%%%%%%%% Fig. 5 %%%%%%%%%%%%%%%%%%%%%%%%%%%%%%%%%
%%%%%%%%%%%%%%%%%%%%%%%%%%%%%%%%%%%%%%%%%%%%%%%%%%%%%%%%%%%%%%%%%%%%%

\section{Conclusions}

We have investigated the electrostatic properties of an electrolyte confined between charged surfaces in contact with a bulk salt reservoir.   A DFT approach combined with a bulk-HNC expansion was employed to calculate the density profiles and the forces acting on the surfaces. Special attention was paid to the local charge neutrality violation in a confined electrolyte. Contrary to the traditional Donnan approach -- in which electroneutrality is enforced by the introduction of a potential difference across the system boundaries -- in our calculations charge neutrality has not been assumed {\it a priori}. The model system can be used to describe the situation in which an electrolyte is confined in carbon nanoporous, for which experimental evidences of local electroneutrality violation has been recently reported \cite{Wu15}. The breakdown of electroneutrality occurs naturally when confined electrolyte is able to exchange particles with a bulk reservoir. Furthermore, the net charge within the confined region can be 
controlled by electrolyte properties other than ionic specificity, such as the  salt concentration and charge asymmetry, as well as the surface charge of the confining walls. In particular, we find that the degree of charge neutrality violation is much more pronounced in the limit of small ionic strengths. These results are in line with the ones obtained previously using a three-point extended HNC-MSA approach \cite{Loz96_1,Loz96_2}.  The results of the present theory were compared to simulations based on a recently introduced efficient implementation of Ewald summation method in a slab geometry, Ref [\onlinecite{San16}].  The agreement between the theory and simulations is excellent.

The system under investigation can also be used to study interactions between colloidal particles \cite{Ohshi74_1,Ohshi74_2,Ohshi74_3,Ohshi06,Pai09}. 
In the traditional Derjaguin, Landau, Verwey, and Overbeek (DLVO) theory it is assumed
that charge neutrality is satisfied in the inter-particle region. On the other hand, the discussion in
the present paper shows that at short separations between the colloidal surfaces charge neutrality is
violated. In order to asses the effect of charge neutrality violation, we have used the contact value
theorem to calculate the force between charged planar surfaces inside an electrolyte
solution. We find that at large salt concentrations, and in some range of wall separations, the net
force on each surface is attractive in the case of multivalent electrolytes. Using Derjaguin approximation,
it should now be possible to construct the effective interaction
potential between charged colloidal particles of radius $R$. The work along these lines is currently in progress.

\section{Acknowledgments}
This work was partially supported by the CNPq, FAPERGS, CAPES, INCT-FCx, and by the US-AFOSR under the grant 
FA9550-12-1-0438.

\appendix

\section{Fundamental Measure Theory}

Here we discuss briefly the FMT. The basic assumption of the FMT is that the excess free energy of a hard sphere fluid has the general form
\begin{equation}
\beta\cF^{\rex}=\int \beta\Phi[\{n_{\alpha}(\mr)\}]d\mr,
\label{A11}
\end{equation} 
where the excess free energy density $\Phi[\{n_{\alpha}(\mr)\}]$ is a {\it local} functional of weighted densities $\{n_{\alpha}(\mr)\}$, which are in turn defined as:
\begin{equation}
n_{\alpha}(\mr)=\sum_{i}\int\rho_{i}(\mr')w_{i}^{(\alpha)}(\mr-\mr')d\mr'.
\label{A12}
\end{equation} 

The simple form of Eq. (\ref{A11}) is strongly suggested by the leading term of $\cF[\{n_{\alpha}(\mr)\}]$ in its low-density diagrammatic expansion. In this limit, the set of weight functions $w_{i}^{(\alpha)}(\mr)$ can be inferred from the deconvolution of the underlying Mayer functions $f_{ij}(\mr)$. It is composed of the four scalar functions,
\begin{eqnarray}
w_{i}^{(3)}(\mr) & = & \Theta(a_{i}-r)  \nonumber \\
w_{i}^{(2)}(\mr) & = & \delta(a_{i}-r)  \nonumber \\
w_{i}^{(0)}(\mr) & = & \dfrac{w_{i}^{(2)}(\mr)}{4\pi a_{i}^2},
\label{A13}
\end{eqnarray} 
together with two vector weight functions $\mathbf{w}_{i}^{(\alpha)}$,
\begin{eqnarray}
\mathbf{w}_{i}^{(2)}(\mr) & = & \delta(a_{i}-r)\dfrac{\mr}{r}  \nonumber \\
\mathbf{w}_{i}^{(1)}(\mr) & = & \dfrac{\mathbf{w}_{i}^{(2)}(\mr)}{4\pi a_{i}}.  
\label{A13}
\end{eqnarray} 
In the context of the Scaled-Particle Theory (SPT) these functions characterize the fundamental measures of hard spheres \cite{Ros88,Ros02}. Note that the corresponding weighted functions in (\ref{A12}) have dimensions of $[\text{length}]^{3-\alpha}$. Since the free energy density $\beta\Phi$ has clearly dimensions of $[\text{length}]^{-3}$, it follows that a it can be quite generally written as the following combination of weighted densities \cite{Ros90}:
\begin{equation}
\beta\Phi[\{n_{\alpha}\}]=f_1 n_{0}+f_2 n_{1}n_{2}+f_3 \mathbf{n}_{1}\cdot\mathbf{n}_{2}+f_4 n_2^{3}+f_5 n_2|\mathbf{n}_2|^2, 
\label{A14}
\end{equation}
with the coefficients $f_i$ being all functions of the dimensionless weighted density $n_3$. The next step in the FMT approach is to impose some conditions to be fulfilled in the limit of homogeneous density distribution, leading to a differential equation from which the coefficients $f_i$ can be determined. The condition used determine the version of FMT. In Rosenfeld's original formulation, a SPT condition is applied in the limit where $a_i\rightarrow \infty$, resulting in the Percus-Yevick (PY) compressibility equation of state for the bulk fluid. The White-Bear version, on the other hand, requires that the (more accurate) Mansoori-Carnahan-Starling-Leland (MCSL) equation of state for hard-sphere mixtures be recovered in the bulk limit \cite{Roth02,Yu02,Man71}. The resulting White-Bear excess free energy density reads:
\begin{equation}
\beta\Phi[\{n_{\alpha}\}]=-n_0\ln(\gamma_3)+\dfrac{n_1 n_2-\mathbf{n}_1\cdot\mathbf{n}_2}{\gamma_3}+\dfrac{n_2(n_{2}^2-3|\mathbf{n}_2|^2)}{36\pi\gamma_{3}^{2}n_3^2}[n_3+\gamma_{3}^{2}\ln(\gamma_3)],
\label{A15}
\end{equation}
where $\gamma_3\equiv 1-n_3$. For a given set of density profiles $\{\rho_{i}(\mr)\}$, the White-Bear hard-sphere excess free energy follows then from the direct applications of Eqs. (\ref{A11}), (\ref{A12}) and (\ref{A15}). We finally note that the corresponding hard-sphere excess single-particle correlations used in Eq. (\ref{dist3}) are given by:
\begin{equation}
c_{i}^{\rhc}(\mr)=-\sum_{\alpha}\int\dfrac{\partial \beta \Phi}{\partial n_{\alpha}}\biggr\arrowvert_{n_{\alpha}(\mr')}w_{i}^{(\alpha)}(\mr'-\mr)d\mr'.
\label{A16}
\end{equation}

\section{Thermodynamic route for the wall forces}
         
We now show that the forces between the walls given by Eq. (\ref{F2}) also follows from the direct thermodynamic calculation of the wall osmotic pressure, regardless of the underlying approximations invoked for the free energy calculation. The osmotic pressure on the walls separated by a distance $d$ can be written as:
\begin{equation}
\beta \Pi=-\dfrac{\partial \beta\Omega}{\partial d}.
\label{A21}
\end{equation}
According to Eq. (\ref{grand}), the grand canonical potential $\Omega$ for the two plate system is: 
\begin{equation}
\beta \Omega=\beta\cF+\sum_{i}\int_{-L_T/2}^{L_T/2}\rho_{i}(z)[\beta\mu_i+\beta\phi_i(z)]dz+2\pi\lambda_B d\sigma^2,
\label{A22}
\end{equation}
where $L_T\equiv L_{z}/4$ represents the system size across the $z$ direction (see Fig.~\ref{fig:fig1}). The last term on the right-hand side is the electrostatic energy resulting from the direct wall-wall interaction. It has to be included since the remaining two terms only contain ion-ion and wall-ion interactions. The resulting osmotic pressure is:
\begin{equation}
\beta\Pi=-\beta\dfrac{\partial\cF[\{\rho_{i}(z)\}]}{\partial d}-\dfrac{\partial}{\partial d}\left[\sum_{i}\int_{-L_{T}/2}^{L_{T}/2}\rho_{i}(z)\left(\beta\mu_i+\beta\phi_i(z)\right)dz\right]-2\pi\lambda_B \sigma^2.
\label{A23}
\end{equation}
Now, since the intrinsic free energy $\cF[\{\rho_{i}(z)\}]$ only depends on the ionic interactions, it can not have any explicit dependence on the particular external potential the particles are subjected to, and has therefore no explicit dependence on $d$. All changes in $\cF[\{\rho_{i}(z)\}]$ come exclusively from the corresponding changes in the ionic profiles as $d$ is varied. Applying the usual rule for the derivative of a functional, we find:
\begin{equation}
\beta\dfrac{\partial\cF[\{\rho_{i}(z)\}]}{\partial d}=\beta\sum_{i}\int_{-L_T/2}^{L_T/2}\dfrac{\delta\cF}{\delta\rho_{i}(z)}\dfrac{\partial \rho_{i}(z)}{\partial d}dz=-\sum_{i}\int_{-L_T/2}^{L_T/2}[\beta\mu_i+\beta\phi_{i}(z)]\dfrac{\partial \rho_{i}(z)}{\partial d}dz.
\label{A24}
\end{equation}
In the last equality, the Euler-Lagrange equilibrium condition $\delta\cF/\delta\rho_i=-(\mu_i+\phi_i(z))$ has been employed. With the above result, Eq. (\ref{A23}) for the osmotic pressure can be simplified to:
\begin{equation}
\beta\Pi+2\pi\lambda_B \sigma^2=-\sum_{i}\int_{-L_{T}/2}^{L_{T}/2}\rho_{i}(z)\dfrac{\partial}{\partial d}[\beta\mu_i+\beta\phi_{i}(z)]dz.
\label{A25}
\end{equation}
Since the chemical potentials $\mu_{i}$ have the purpose of fixing the ionic bulk concentrations $\bar{\rho}_{i}$ -- and these are in the present formulation kept constant as $d$ changes -- the first derivative on the right-hand side can be set to zero. We further notice that the wall-ion interaction potential can be split into hard-core and electrostatic contributions, $\phi_i(z)=\phi_{i}^{\rcoul}(z)+\phi_{i}^{\rhc}(z)=\beta q_{i}\phi_{p}^{\rcoul}(z)+\phi_{i}^{\rhc}(z)$, which are given by Eqs. (\ref{elec}) and (\ref{phi_hc}), respectively. It follows that the osmotic pressure can be written as:
\begin{equation}
\beta\Pi+2\pi\lambda_B \sigma^2=-\sum_{i}\int_{-L_{T}/2}^{L_{T}/2}\rho_{i}(z)\left[\dfrac{\partial\beta\phi_{i}^{\rhc}(z)}{\partial d}+\dfrac{\partial \beta\phi_{i}^{\rcoul}(z)}{\partial d}\right]dz\equiv \beta\Pi^{\rhc}+\beta\Pi^{\rcoul}.
\label{A26}
\end{equation}   
We now turn to the calculation of each contribution on the right-hand side of (\ref{A26}) separately. First, it is convenient to rewrite the first integral as:
\begin{equation}
\beta \Pi^{\rhc}=-\sum_{i}\int_{-L_T/2}^{L_T/2}\rho_{i}(z)\dfrac{\partial\beta\phi_{i}^{\rhc}(z)}{\partial d}dz=\sum_{i}\int_{-L_T/2}^{L_T/2}\rho_{i}(z)e^{\beta\phi_{i}^{\rhc}(z)}\dfrac{\partial}{\partial d}e^{-\beta\phi_{i}^{\rhc}(z)}dz.
\label{A27}
\end{equation}   
The so-called cavity functions $e^{-\beta\phi_{i}^{\rhc}(z)}$ are either zero at particle overlap with the hard walls or one otherwise. Similarly, the function $e^{\beta\phi_{i}^{\rhc}(z)}$ goes to infinity at ion-wall overlap and to unity anywhere else. It follows from Eq. (\ref{phi_hc}) that 
\begin{equation}
e^{-\beta\phi_{i}^{\rhc}(z)}=\Theta(-z-d/2-a_i)+\Theta(z+d/2-a_i)-\Theta(z-d/2+a_i)+\Theta(z-d/2-a_i). 
\label{A27_b}
\end{equation}
Performing the straightforward differentiation with respect to $d$ provides:
\begin{equation}
\dfrac{\partial}{\partial d}e^{-\beta\phi_{i}^{\rhc}(z)}=\dfrac{1}{2}\left[\delta(z+d/2-a_i)-\delta(-z-d/2-a_i)+\delta(z-d/2+a_i)-\delta(z-d/2-a_i)\right].
\label{A28}
\end{equation} 
Using the above result in Eq. (\ref{A27}) and performing the integration, we find:
\begin{equation}
\beta\Pi^{\rhc}=\dfrac{1}{2}\sum_{i}[\rho_{i}(-d/2+a_i)-\rho_{i}(-d/2-a_i)+\rho_{i}(d/2-a_i)-\rho_{i}(d/2+a_{i})]=\sum_{i}\rho_{i}(-d/2+a_i)-\rho_{i}(-d/2-a_i),
\label{A29}
\end{equation}
where in the last equality the parity of the distribution functions $\rho_{i}(z)=\rho_{i}(-z)$ was used. As for the electrostatic contributions $\Pi^{\rcoul}$ in Eq. (\ref{A26}), we first notice that Eq. (\ref{elec}) can be used to write the ion-wall electrostatic potential as:
\begin{equation}
\beta\phi_{i}^{\rcoul}=4\pi\lambda_{B}\sigma \alpha_{i}\left[(z+d/2)\Theta(-z-d/2)-(z-d/2)\Theta(z-d/2)\right].
\label{A210}
\end{equation}
Differentiation with respect to $d$ provides:
\begin{equation}
\beta\dfrac{\partial\phi_{i}^{\rcoul}(z)}{\partial d}=2\pi\lambda_{B}\sigma \alpha_{i}\left[\Theta(-z-d/2)+\Theta(z-d/2)\right].
\label{A211}
\end{equation}
The electrostatic contribution to the osmotic pressure then becomes:
\begin{equation}
\beta\Pi^{\rcoul}=-2\pi\lambda_{B}\sigma \left[\sum_{i} \alpha_{i}\int_{-L_T/2}^{-d/2}\rho_{i}(z)dz+\int_{d/2}^{L_T/2}\rho_{i}(z)dz\right].
\label{A212}
\end{equation}
Now, using the overall electroneutrality condition for the system as a whole, Eq. (\ref{total_cn}), the term in parenthesis can be clearly identified as $-2\sigma-\sigma_{\rin}$, and the above relation can be simplified to:
\begin{equation}
\beta\Pi^{\rcoul}=2\pi\lambda_{B}\sigma(2\sigma+\sigma_{\rin}).
\label{A213}
\end{equation}
Substitution of Eqs. (\ref{A29}) and (\ref{A213}) in Eq. (\ref{A26}) leads to osmotic pressure across the charged walls:
\begin{equation}
\beta\Pi=2\pi\lambda_{B}\sigma(\sigma+\sigma_{\rin})+\sum_{i}\rho_{i}(-d/2+a_i)-\rho_{i}(-d/2-a_i),
\label{A213}
\end{equation}
which precisely recovers the result (\ref{F2}) for the force per unit of area exerted on the wall at $z=-d/2$. It is important to emphasize that no assumption has been made on $\cF[\{\rho_{i}(z)\}]$ other than its fulfillment the Euler-Lagrange equilibrium condition. The result above is, therefore, quite general and independent on the particular set of approximations employed in the construction of $\cF^{\rex}[\{\rho_{i}(z)\}]$. This is to be contrasted with the situation in which electrolytes are present at only one single side of the surface, where then the contact condition does depend on the particular free energy functional \cite{Fry12}. When the charged surface lies in between two electrolytes, the mutual interaction between these electrolytes will also play a role on the net force acting upon the surface \cite{Loz96}. The numerical derivative in Eq. (\ref{A21}) can in practice be performed as an accuracy check of the calculated ionic density profiles in Eq. (\ref{A213}), since the corresponding values are usually sensitive to numerical precision \cite{Wan11_2}.

%\bibliography{ref}


\end{document}